\documentclass[journal,comsoc]{IEEEtran}
\usepackage[T1]{fontenc}

\ifCLASSINFOpdf
\else
\fi
\usepackage{amsmath}
\usepackage[cmintegrals]{newtxmath}
\usepackage{array}
\usepackage{multirow}
\usepackage{longtable}
\usepackage[english]{babel}
\usepackage{blindtext}
\usepackage{xspace}
\usepackage{rotating}
\usepackage{subcaption}
\usepackage[hyphens]{url}
\usepackage{cleveref}
\crefformat{section}{\S#2#1#3} 
\crefformat{subsection}{\S#2#1#3}
\crefformat{subsubsection}{\S#2#1#3}  

\newcommand{\dc}{data center\xspace}
\newcommand{\dcs}{data centers\xspace}

\newcommand{\eg}{\textit{e}.\textit{g}.}

\newcommand{\thss}{\textsuperscript{th}}
\newcommand{\millis}{$\mathrm{ms}$\xspace}

\def\etal{{\it et al.}}
\newcommand{\tabitem}{\textbullet~}

\begin{document}
%
\title{Measurement-based Resource Allocation and Control in Data Centers: A Survey}
%
%
%

\author{Diana Andreea Popescu, {\emph{University of Cambridge}}
}

%
%

\markboth{D.A. Popescu, \emph{Measurement-based Resource Allocation and Control in Data Centers: A Survey}}%
{}
%



\maketitle
\begin{abstract}
Data centers have become ubiquitous for today's businesses. 
From banks to startups, they rely on cloud infrastructure
to deploy user applications. In this context, it is vital
to provide users with application performance guarantees.
Network interference is one of the causes of unpredictable 
application performance, and many solutions have been proposed
over the years.
The main objective of this survey is to familiarize the reader
with research into network measurement-based resource
allocation and control in \dcs, focusing on network resources
in order to provide cloud performance guarantees. 
We start with a primer on general network measurement techniques
and \dc network and applications to give the reader context.
We then summarize the characteristics of
network traffic and cluster workloads in \dcs, which are pivotal 
for measurement-based allocation and control.
We study and compare network monitoring in \dcs,
giving an overview on their evolution from Software-Defined Networking (SDN) to programmable dataplanes-based.
The network monitoring information can serve as input to cluster allocation and scheduling decisions.
We next categorize cluster scheduling frameworks,
and perform an analysis of those that provide
network guarantees in \dcs, and we also look at emergent Machine Learning-driven
resource allocation and control. 
We conclude with a discussion about future research directions.

\end{abstract}
\section{Introduction}

Networks represent an important component of modern computing, shaping an interconnected world.  
While high-level applications, such as games, music and video streaming, office tools, play an important role in people's lifes, the importance of communication between computers cannot be overstated.  Without a fast and reliable means of communication, these applications would be running only on the local computers with no external input, safe for the information obtained through compact disks or memory sticks. The Internet has evolved over the years into such a fast and reliable means of communication. At the heart of this evolution have been the strong need of people to communicate with each other, and the information that people want to disseminate to the world. As a result, a person can contact anyone over the Internet through electronic mail, voice or video call, and information on just about anything is nowadays easily available worldwide to anyone with an Internet connection. These two key needs have fueled the development of networks over the years, making the Internet ubiquitous. 

Services such as web search, social networking, online shopping, content streaming, instant messaging, video calling, digitised newspapers, books, or research articles, form the online landscape today. Because of the huge number of people that access these services, enormous computing resources are needed.
This demand has led to the development of specialised warehouse-scale computers (WSCs)~\cite{warehouse} that are housed in \emph{data centres}. Data centres contain many hundreds of thousands of such computers, powering the above-mentioned services and many more. But designing and operating such complex systems require expert knowledge. As a result, only a few companies, such as Amazon, Google, IBM, Oracle or Microsoft, develop and manage their own \dcs. This complexity has given rise to \emph{cloud computing}: businesses rent compute, storage and network resources from specialised companies, \emph{cloud providers}, to power their services, instead of developing and maintaining their own infrastructure. Naturally, businesses require and expect predictability from the rented resources, which translates into \emph{predictable performance} for their applications. The customer, or \emph{tenant}, expectations are encoded into contracts called \emph{Service Level Agreements (SLAs)}, with specific objectives, \emph{Service Level Objectives (SLOs)}, defined in collaboration with the cloud provider. The objectives refer to quantifiable metrics, such as availability, throughput, or response time~\cite{diana-report}.

The infrastructure in \dcs is shared amongst different tenants, giving rise to possible interference between different applications, which in turn can lead to unpredictable application performance~\cite{diana-thesis}. Interference can appear in multiple places in the \dc: at the servers (\emph{hosts} or \emph{end-hosts}), where tenant applications that run inside \emph{virtual machines (VMs)} share the underlying server hardware, or in the network, with some tenants sending traffic that causes packets to queue behind each other in switches, increasing the packets' \emph{network latency}, or \emph{network delay}, or \emph{one-way delay (OWD)} (the time a packet takes to travel from the source to the destination across the network)~\cite{mogul:2015, Barroso}. 
Several approaches to minimise interference have been proposed, both at the end-host and in-network. Avoiding or reducing the interference at the end-host can be achieved through thoughtful \emph{cluster scheduling}: avoiding the colocation of applications that compete for the same end-host resources by placing these applications on different hosts that meet their resource requirements. In the network, interference effects can be avoided or reduced through \emph{flow scheduling}~\cite{pfabric, fastpass, phost, qjump, ndp} and traffic \emph{load balancing}~\cite{Hedera, MicroTE, Conga}. Pinpointing the exact cause of the interference is as challenging as \dc systems are complex~\cite{vnet-pingmesh}. 

This survey is organized as follows. Section~\ref{sec:related} discusses related survey and defines the scope for our survey. Section~\ref{sec:network-measurement} presents general network measurement definitions and techniques. Section~\ref{sec:dc-network} gives an overview \dc network architecture, and presents the type of applications that run in \dcs. Section~\ref{ss:dc-characteristics} summarizes the main findings of several studies on \dc network traffic characteristics. Section~\ref{sec:cluster-workloads} presents the cluster workloads characteristics. Section~\ref{sec:network-monitoring} discusses the evolution of network monitoring systems for \dcs, from SDN to programmable data planes. Section~\ref{sec:bk-cluster-scheduler} presents conventional cluster schedulers, network-aware cluster schedulers and emergent ML-driven ones. We conclude in Section~\ref{sec:future} and present future research directions for each area discussed in our survey.

\section{Related surveys and scope of this survey}
\label{sec:related}

Several works dealt with issues related \dc networking. \cite{sdc-energy} surveys models for energy consumption in \dcs.
\cite{sdc-virt} discusses virtualization in \dcs.
\cite{sdc-lb} surveys load balancing mechanisms in \dcs.
\cite{sdc-tc} presents proposals for traffic control in \dcs:
traffic management, transmission control, traffic shaping, prioritization, 
load balancing, multipathing, and traffic scheduling. \cite{sdc-io}
discusses \dc network infrastructure and operations. \cite{sdc-opt}
surveys resource management techniques and categories depending on the
resource granularity: virtual machines, physical machines and
applicatios. \cite{sdc-cluster} presents cluster frameworks for scheduling
in \dcs at the application, network and switch level. 

Our survey presents a network measurement-based perspective on resource allocation
and control in \dcs, and it distinguishes itself from earlier surveys from the following perspectives:
\begin{itemize}
\item We focus on network measurement and how measurement systems are used for 
resource allocation and control mechanisms.
\item We discuss the state-of-the-art across different areas (network measurement,
\dcs network and applications, network monitoring systems, cluster scheduling) to highlight
the synergies between these areas and show how these areas fit together
to improve application performance in \dcs by leveraging network monitoring information.
\item We analyze the literature and we identify gaps across different areas, which
can only be solved by taking a holistic approach across the \dc management stack. 
\end{itemize}

\section{Network measurement 101}
\label{sec:network-measurement}

\subsection{Network measurement definitions}

Before looking into measurement techniques, we give a brief overview of traffic properties which are usually measured, as presented in \cite{Crovella}. 
\emph{Packet delay} is an additive metric and is the sum of routing delay (time spent inside a router), transmission delay (time needed to put a packet onto a link) and propagation delay (time needed by a packet to traverse the link from one end to another). Further, the routing delay can be decomposed into packet processing delay (time to determine the output port for the packet), queueing delay (time spent waiting in router's output queues) and other additional delay, such as marshalling and unmarshalling. The queueing delay can be seen as a measure of congestion on the output link. 
Another property is the rate of \emph{packet loss}: the phenomenon which occurs when a network device drops the packet due to congestion, or because the packet is identified corrupted and then dropped. \emph{Throughput} is the rate at which traffic flows through the network and is measured in bits per time unit. \emph{Packet jitter} represents the variability of packet interarrival time. An important property for application performance is \emph{goodput}, which is the rate at which the application endpoint successfully receives data. 
There are other properties that can be used to characterise the traffic: time series of byte counts (to quantify the workload represented by traffic), and the distribution of packet sizes encountered. 
Another property is the \emph{ON/OFF activity} in network traffic (ON state represents activity, OFF state means silence). The activity can be viewed at three different levels: the packets themselves are an ON/OFF process, packets form \emph{trains} from a source to a destination (defined by a given interarrival threshold), and a collection of trains form a \emph{session}. 
\emph{A flow} can be defined as a set of packets having the same predefined properties, and that are exchanged between two endpoints.

There are two types of measurement methods: \emph{active} and \emph{passive} methods \cite{Crovella}. Active measurement methods imply injecting additional packets into the network and observing their behaviour. The best known active measurement tools are \emph{ping} and \emph{traceroute} (see Table~\ref{tab:basic-network-monitoring-tools}). Network latency from a
source host to a destination host is often measured as Round-Trip Time (RTT) using \emph{ping}.
A disadvantage of active methods is that they create additional network load and can thus bias the measurements. On the other hand, passive measurement methods rely solely on observing the traffic without generating additional traffic.

\subsection{Network monitoring techniques and tools}
\label{tools}

Table~\ref{tab:basic-network-monitoring-tools} presents classical network monitoring techniques and tools, and the measurements they offer. One of the most well known method to get network statistics is through the Simple Network Management Protocol (SNMP) counters~\cite{snmp}. The information that can be obtained by polling these counters regularly is the number of bytes received and sent on each interface of the network device or the number of packets received and sent on each interface.
Although maintaining these counters does not have a significant performance overhead on the network device, they offer only a course-grained view of the network traffic, because they are limited by the polling interval (typically 300s), and by imprecise timestamps. The timestamp is supplied externally, and it is limited in numerical range, leading to the need to detect roll over. SNMP's resolution is sufficient for low throughput traffic.

\emph{Packet sampling}, as used in NetFlow~\cite{netflow}, sFlow~\cite{sflow}, PSAMP~\cite{psamp}, entails capturing only a subset of packets to reduce the number of records~\cite{Crovella}. Packet sampling can be done using a constant or variable sampling rate. In the case of constant sampling, there are several possibilities: i) random sampling (packets are sampled with a fixed probability $0<p<1$); ii) deterministic sampling (packets are sampled periodically, meaning every $N$th packet is sampled); iii) stratified sampling (packets are first divided into subsets, and then sampling is applied within each subset).
Another type of sampling is \emph{trajectory sampling}~\cite{trajectory, Crovella}. In this technique, if a packet is selected for sampling at some device in the network, then it will be selected at all the other devices in the network. Amongst the uses of trajectory sampling are obtaining packet delays (important for SLAs), or tracing denial of service attacks. 

NetFlow~\cite{netflow} is a standard introduced by Cisco to monitor IP flows, which are usually identified by the $5$ tuple (source and destination IP address, source and destination port and protocol number). The active TCP and UDP flows are kept in a cache. When a packet is received at the switch, NetFlow checks to see if the packet pertains to a cached flow by matching on the header fields. If it does, then the associated flow counters are incremented, and if it does not, a new flow entry is created in the cache. For deciding when to send flow records to the collector for analysis, several policies can be configured: i) on a TCP flow completion, which can be detected when seeing a packet with a FIN or RST flag; ii) when a flow has been idle for a configured timeout; iii) if a hard timeout is configured; iv) when the cache is full and an entry must be evicted. Timeouts can be specified at granularity of seconds. 
Sampled NetFlow~\cite{samplednetflow} can sample $1$ in $N$ consecutive packets that traverse the switch. 

sFlow~\cite{sflow} is a standard implemented in most new switches to provide packet sampling and port counter sampling. For packet sampling, a switch selects $1$ in $N$ packets on each input port. The sampled packets' headers are forwarded to a collector along with metadata that includes the sampling rate, the switch ID, the timestamp of capture, the input and output port numbers.  

Port mirroring~\cite{port-mirroring} involves copying all packets seen on a switch port to a different switch port for further analysis.

\begin{table*}
\begin{center}
\setlength{\tabcolsep}{3pt}
\small
\begin{tabular}{|l||l|}
\hline
	\textbf{Tool} & \textbf{Measurement}\\
\hline
\hline
    Ping &  RTT; packet loss ratio \\
\hline
    Traceroute & route 'guess'; RTT\\
\hline
    SNMP~\cite{snmp}  & switch port counters\\
\hline
    NetFlow~\cite{netflow} &  flow counters\\
\hline
    sFlow~\cite{sflow} & packet sampling \\
\hline
    iperf~\cite{iperf} & throughput \\
\hline
    Cisco IP SLA~\cite{ipsla} & RTT (average); one-way delay; packet loss \\
\hline
    Port mirroring~\cite{port-mirroring} & copies of all packets from a port\\
\hline
    NTP~\cite{ntp} & round-trip delay\\
\hline
	PTP~\cite{ptp} & master-to-slave delay; slave-to-master delay; estimated one-way delay\\
\hline	
\end{tabular}
\caption{Classical network monitoring tools.}
\label{tab:basic-network-monitoring-tools}
\end{center}
\end{table*}

\section{Data centres 101}
\label{sec:dc-network}

\subsection{Data centre network architecture}
\label{ss:dc-network}

A \dc network architecture is comprised of the topology of the network that interconnects the servers, of the switches deployed in the network, of the end-host network configuration and of the communication protocols used. Companies do not reveal full details of their \dc architectures, since the performance given by their infrastructure can be an advantage over competitors, especially in the cloud computing business.

\textbf{Network hardware and topology} The most common topology used for \dcs is \emph{fat tree}~\cite{fattree}, which is based on Clos networks~\cite{clos}. Clos networks, originally designed for telephone circuit switches, are multi-stage circuit-switching networks, with three stages: the ingress stage, the middle stage, and the egress stage. Clos networks are strict-sense non-blocking networks, meaning that any input can be connected to an unused output without having to rearrange existing connections. Fat trees, on the other hand, are rearrangeable non-blocking networks, meaning that with a certain arrangement of the connections, any input can be connected with any unused output.
An important feature of a network topology is the \emph{bisection bandwidth}.
The bisection bandwidth of a network is the bandwidth available between the two partitions when the network is partitioned in half. Full bisection bandwidth means that any input can communicate with any unused output at \emph{full line-rate}. Non-blocking networks provide \emph{full bisection bandwidth}. The edge of the network is usually \emph{oversubscribed}.
Fat tree topologies can offer full bisection bandwidth, but it may be difficult to achieve this while also avoiding packet reordering in TCP flows~\cite{fattree}. 

\begin{figure}[!h]
\centering
\includegraphics[width=0.5\textwidth]{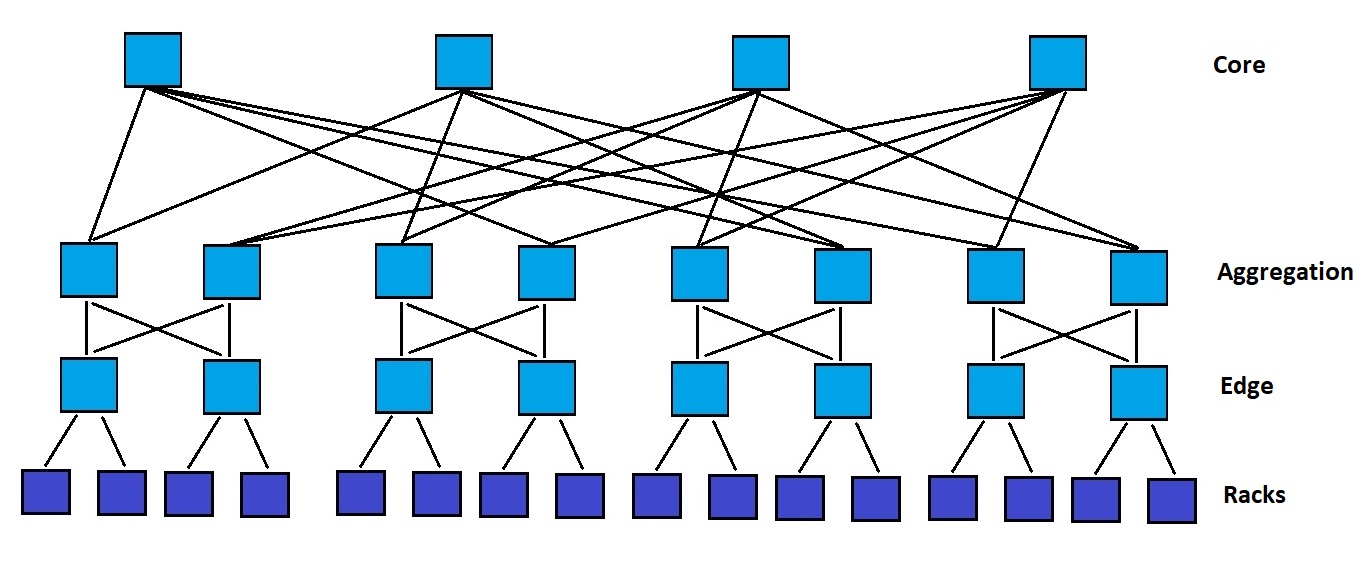}\caption{A \dc fat-tree topology.}\label{fig:fat-tree}
\end{figure}  

Google~\cite{singh2015jupiter}, Microsoft~\cite{vl2} and Facebook~\cite{facebook-dc} \dc network topologies follow the fat tree design. A $k$-ary fat tree, shown in Figure~\ref{fig:fat-tree}, has $3$ layers of $k$-port switches: core (the top layer), aggregation (the middle layer) and edge (the lowest layer, which is connected to the host layer). The hosts are grouped into \emph{racks}, and are connected to edge switches, called \emph{Top-of-Rack (ToR)} switches.
The fat tree has $k$ \emph{pods}, and a pod contains two layers of switches, each layer having $\frac{k}{2}$ switches, and $\frac{k^2}{4}$ hosts. A $k$-port switch from the edge layer is connected to $\frac{k}{2}$ hosts and to $\frac{k}{2}$ switches from the aggregation layer. The fat tree topology has $(\frac{k}{2})^{2}$ core switches. A $k$-port core switch is connected to every pod, having its $i$-th port connected to the $i$-th pod such that an aggregation switch is connected to $\frac{k}{2}$ consecutive core switches. The fat tree topology provides multiple equal cost paths between every two hosts. Switches use Equal Cost Multipath routing (ECMP)~\cite{ecmp} to decide on which of the equal cost paths a flow should be sent. ECMP hashes the 5-tuple (IP source and destination address, source and destination port and protocol ID) of a packet, and based on the hash determines which path the flow will take. 

The Google~\cite{singh2015jupiter} network has evolved over the years. Google develops their own switches using $16 \times 40$Gb/s merchant silicon. In the Jupiter network, a ToR switch has $48 \times 40$Gb/s connections to hosts in the rack and $16 \times 40$Gb/s to the aggregation switches. Four such switches form a Middle Block (MB), which serves as building block in the aggregation
block. The logical topology of an MB is a 2-stage blocking network, with $256 \times 10$Gb/s connections to ToRs
and $64 \times 40$Gb/s connections to the spine. Each ToR connects to eight MBs with dual redundant $10$Gb/s links. An aggregation block has $512 \times 40$Gb/s or $256 \times 40$Gb/s links towards the spine blocks. A spine block has six switches with $128 \time 40$Gb/s ports to the aggregation blocks. There are 64 aggregation blocks.

The Facebook \dc architecture~\cite{facebook-dc} has four planes of spine (core) switches. Each plane can accommodate up to $48$ spine switches. Each fabric (aggregation) switch of each pod connects to each spine switch within its plane. A pod has 48 server racks, with each pod being served by four fabric switches, one from every plane. A rack can have up to $192$ hosts. Each ToR has $4\times 40$Gbs/s uplinks. End-host have 10Gb/s connections. There is a 4:1 fabric oversubscription from rack to rack, with $12$ spine switches per plane. Facebook also develops their own switches.

VL2~\cite{vl2} is a 3-tier architecture where the core tier and the aggregation tier form a folded Clos topology. Other proposed data center network designs include Dcell \cite{Dcell}, BCube \cite{BCube}, CamCube \cite{CamCube}, Jellyfish \cite{Jellyfish}, Xpander~\cite{xpander}. BCube \cite{BCube} is a \dc network architecture based on a hyper-cube topology. Jellyfish \cite{Jellyfish} is a random graph topology designed to support easy incremental expansion of the \dc.  

\emph{Hardware resource disaggregation} is an emerging trend that will see the traditional rack replaced by pools of different resources (CPU, DRAM, disk)~\cite{gao2016,legoos, intel-disaggregated} communicating over a high-speed network. This architecture presents several benefits: improved resource utilisation, failure isolation, and flexibility in adding or removing resources. 

\textbf{Network protocols}
The \dc protocol stack is based on the traditional TCP/IP stack, but it has evolved to deal with the challenges inherent to this type of environment: scale, cost, competing demands of applications (high throughput vs. low latency), unpredictable network traffic patterns. Most of the \dc operators today are using IP version 4 (IPv4) within their networks. Facebook uses only IP version 6 (IPv6) in their internal networks.

Traditional broadcast mechanisms such as Address Resolution Protocol (ARP)~\cite{arp} do not scale in \dcs~\cite{vl2, portland}. To solve this issue, \dcs are managed via IP layer routing protocols, such as the Border Gateway Protocol (BGP)~\cite{bgp}. For example, Facebook uses BGP4, which is complemented by a centralised BGP controller that is able to override routing paths. Google developed its own protocol, named FirePath~\cite{singh2015jupiter}, which is a custom Interior Gateway Protocol (IGP). FirePath implements a centralised topology state distribution, and a distributed forwarding table computation.
Microsoft's \dc protocol is thought to be similar to that of VL2~\cite{vl2}: traffic originating from the edge switches is forwarded first to a randomly selected intermediate switch and then to the actual destination. 

TCP's congestion control algorithm is not optimal for environments which have the characteristics of a \dc network (high-bandwidth, low latency). Consequently, TCP's performance suffers because of different issues, such as incast, bursty packet drops, and large queue buildup~\cite{alizadeh2010dctcp}. Thus, different TCP variants customised for \dcs were developed over the years~\cite{alizadeh2010dctcp, hull, d2tcp}. The first such transport protocol variant was DCTCP~\cite{alizadeh2010dctcp}. DCTCP signals queue buildup earlier on through the use of the Explicit Congestion Notification (ECN) feature supported by certain switches. Source end-hosts estimate the fraction of packets marked through ECN, and deduce the amount of congestion. Google's \dcs run a variant of DCTCP~\cite{singh2015jupiter}. 

Moreover, clean slate designs have been proposed in this space~\cite{fastpass, ndp}. Fastpass~\cite{fastpass} is a centralised packet scheduler that aims to reduce in-network queueing. It proposes a timeslot allocation algorithm to determine when each packet is sent, along with a path assignment algorithm for each packet. Fastpass moves the queueing at end-hosts. NDP~\cite{ndp} is a radically different approach that requires a new end-host stack and new switches. It uses switches with small buffers. The senders send a full window from the start, with no initial handshake. When congestion occurs, the switches trim packets, removing their payload. The headers are sent to notify the receiver of which senders wish to send to it. Then, the receivers pull data from the senders who want to send to them, since the receivers have received the headers and know from whom to expect data.

\subsection{Data centre applications}
\label{ss:dc-apps}

A key role in the \dc ecosystem is played by the applications that produce network traffic. Due to the scale of the input data and of the user demands, \dc applications are distributed. Data centre applications can be split in the following categories:
\begin{itemize} 
	\item Control and management applications: clock synchronisation (\eg, Precision Time Protocol daemon~\cite{ptpd}), consensus and locking (\eg, Chubby~\cite{chubby}), and cluster management (\eg, Borg~\cite{borg}).
	\item Data storage and retrieval: distributed file systems (\eg, GFS \cite{GFS}), distributed database systems (\eg, Spanner~\cite{spanner}), key-value stores (\eg, Memcached \cite{memcached}). 
	\item Applications serving users' needs: data processing frameworks (MapReduce~\cite{MapReduce} style processing, graph processing (\eg, Apache Giraph~\cite{giraph}, Pregel~\cite{pregel}), stream processing (\eg, Apache Storm~\cite{storm}), machine learning analytics (\eg, Tensorflow~\cite{tensorflow}), Web traffic~\cite{apache-web-server}, search engine and social network backends. In addition to these, tenant applications running in VMs, which can be any of the previous applications or custom applications, have to be mentioned.  
\end{itemize}

There are two aspects of the applications that are important for determining their networking requirements: the communication pattern of the application (which can be mapped to the underlying network) and the properties of representative workloads. 

\textbf{Communication patterns} 
The most common communication patterns~\cite{coflow, Kapoor:2012} are MapReduce~\cite{MapReduce}, partition-aggregate (search engine and social networks backend)~\cite{alizadeh2010dctcp, Kapoor:2012, coflow}, dependent-sequential (constructing a user's home page in a social networking application)~\cite{Kapoor:2012}, star-like (machine learning parameter server~\cite{parameter-server}), and Bulk Synchronous Parallel (BSP) (\eg, Pregel~\cite{pregel}). 
In the \emph{MapReduce} \cite{MapReduce, spark-rdd} pattern, a mapper reads its input from the distributed file system, performs computations on the input read and then writes its intermediate result to disk. A reducer reads the intermediate result from different mappers (the shuffle phase), performs computations on the data, and writes the output to the distributed file system. In the shuffle phase there are $xy$ flows if there are $x$ mappers and $y$ reducers, and at least $y$ flows for writing the final results. 

In the \emph{partition-aggregation} pattern \cite{alizadeh2010dctcp, Kapoor:2012}, in order to provide an answer to a request received from a user, several responses from workers need to be aggregated. The aggregation tree can have multiple levels, with the leafs being the workers and the root being the final aggregator. The \emph{dependent-sequential} pattern \cite{Kapoor:2012} entails that the next request is dependent on the previous request's results. These patterns are common in applications such as Web search and social network content backends.

A \emph{Bulk Synchronous Parallel (BSP)} computation, named superstep, consists of concurrent computation, communication between worker processes, and barrier synchronisation. This pattern is common in graph processing frameworks~\cite{pregel, giraph}. 

\emph{Machine learning (ML) applications} represent a common workload for \dcs~\cite{ml-at-facebook, tensorflow}. In general, an ML application fits a model to input data, and requires multiple iterations until the model's parameters convergence. Due to the huge amount of input data that has to be processed, ML frameworks have a distributed architecture~\cite{parameter-server, spark-rdd, Kim:2016, petuum, tensorflow}. A machine learning framework usually has server nodes that store the globally shared parameters (parameter servers), and worker nodes that do local computations on their part of data or of the model, depending on the chosen approach. In a data parallel approach, the input data is partitioned across the machines and the ML model is shared. In a model parallel approach, the ML model is partitioned across the machines and the input data is shared. In the data parallel approach, each worker node can read and update all the model parameters, while in the model parallel approach, each worker node can access and update only its model parameter partition. In this context, the network plays an important role due to the inherent synchronisation between worker nodes and server nodes to update the model. If the computation is synchronous, after each iteration, the parameter serves aggregate the parameters from the workers. For example, if one of the workers is unreachable over the network or is slow to reply, the overall training time increases due to the wait for that worker's parameter updates. To lessen the importance of communication latency to the completion time of the training, some frameworks~\cite{tensorflow, Kim:2016} use asynchronous communication, or bounded staleness synchronisation~\cite{petuum}, where a certain degree of staleness in the parameters (meaning using parameters from previous iterations) is tolerated, but this can potentially lead to slower converge of the model~\cite{Kim:2016}.

\textbf{Workloads} There are a few studies which analyse the workloads of some data centre applications. One such study is the analysis and modelling of Memcached workloads based on data provided by Facebook \cite{Atikoglu:2012}. The authors describe an analytical model that can be used to generate synthetic workloads whose properties are similar to the real world workloads, and which is implemented in the Mutilate load generator~\cite{Leverich:2014}. MapReduce workloads from Facebook and Cloudera are discussed in \cite{mapreduceworkload}, providing insights about job size, storage access patterns, and cluster load. ML applications usually can be tested with well-known datasets, \eg, MNIST dataset~\cite{mnist} for handwritten digit recognition, ImageNet~\cite{imagenet} dataset for image classification tasks.

\section{Data centre network traffic characteristics}
\label{ss:dc-characteristics}

Data centres network traffic characteristics are seen as sensitive information by companies, as they could reveal details about their network infrastructure to their competitors. As such, there is little information available on this topic. Still, there are three main studies \cite{dctraffic, benson, roy2015} which shed some light on this matter. Additionally, several papers \cite{vl2,alizadeh2010dctcp,infocomvm, orchestra, augmenting} (see Table~\ref{tab:dc-traffic-characteristics} and Table~\ref{tab:dc-traffic-characteristics-2}) present limited measurement studies from different data centres. 

\begin{table*}
\scriptsize
\begin{tabular}{ |p{1cm} ||p{3cm} |p{1.5cm} |p{3cm} |p{2cm} |p{3cm} |}

\hline
\textbf{Study} & \textbf{Data centre} & \textbf{Duration}  & \textbf{Workload} & \textbf{Measurements} & \textbf{Flow sizes} \\ \hline
\hline

\cite{benson}
&
10 data centres 
(3 university \dcs, 
2 private enterprise \dcs, 
5 commercial cloud \dcs)
& 
several weeks
&
Web services, MapReduce, 
file storage, authentication, 
business applications, 
custom software applications, 
email and messaging
&
\tabitem network topology; \newline 
\tabitem packet traces from switches; \newline 
\tabitem SNMP
&
\tabitem 80\% of the flows are smaller than 10KB in size. \newline 
\tabitem 80\% of flows are less than 11 seconds long.

\\
\hline
\cite{dctraffic} &
1500 servers
&
1 PB of measurement data
&
\tabitem MapReduce style jobs; \newline
\tabitem Cosmos distributed file system
&
\tabitem socket level logs; \newline 
\tabitem user application logs;
&

\tabitem most flows are short, 80\% of the flows lasting less than 10s \newline
\tabitem only less than 0.1\% last longer than 200s \newline
\tabitem more than half of the bytes are found in flows that last less than 25s

\\
\hline
\cite{vl2} 
&
1500 node cluster 
&
-
&
data mining on petabytes of data
&
\tabitem SNMP \newline 
\tabitem NetFlow
& 
\tabitem 99\% of flows are smaller than 100MB \newline
\tabitem 90\% of bytes are in flows whose lengths are between 100MB and 1 GB \newline

\\
\hline
\cite{alizadeh2010dctcp}
& 6000 servers in over 150 racks
& 1 month; 150 TB 
& web search and other services
& \tabitem socket level logs; \newline
\tabitem packet level logs; \newline
\tabitem application level logs
& 
\tabitem most background flows are small, but most of the bytes in
background traffic come from large flows \newline

\\
\hline
\cite{augmenting}
&
\tabitem pre-production cluster with O(1K) servers running Dryad (Cosmos) \newline
\tabitem production cluster with O(10K) servers where the web search index is stored and where search results are assembled (IndexSrv)   
&
$76$ hours and $114$ terabytes of data
& 
\tabitem a data mining workload for a large web search engine + jobs which are
a mix of repetitive production scripts (e.g., hourly summaries) and
jobs submitted by users \newline
\tabitem web search (latency sensitive)
& estimate demand matrices and determine hotspot occurrence and predictability \newline
& 
\tabitem medium-sized flows
\\
\hline
\cite{infocomvm} & 
IBM Global Services \newline
\tabitem DC 1: $17000$ VMs \newline
\tabitem DC 2: $68$ VMs
  & $10$ days  
   & - &
latency measurements between each two servers; \newline  
TCP incoming/outgoing connections \newline
&
-

\\
\hline
\cite{roy2015}
&
Facebook data centres 
& 
brief periods of time
&
Web services, MapReduce, 
MySQL, caching
&
\tabitem sampled traffic; \newline 
\tabitem packet traces from switches; \newline 
\tabitem packet traces from host
&
\tabitem Hadoop flows are short; \newline 
\tabitem for other type of applications, flows are long-lived, internally bursty and do not carry a significant number of bytes.
\\
\hline
\end{tabular}
\captionof{table}{Data centre network traffic characteristics - part 1.}
\label{tab:dc-traffic-characteristics}
\end{table*}

Benson~\etal~\cite{benson} present a study of the network traffic characteristics of 10 data centres (3 university \dcs, 2 private enterprise \dcs, 5 commercial cloud \dcs). The data sets used in characterising the network traffic are the following: network topology, packet traces from switches and SNMP polls. The data was collected over several weeks. There are several important findings. Firstly, perhaps predictably, the applications that run in the \dcs depend on the organisation. There is a wide range of applications in each \dc: Web services, MapReduce, file storage, authentication, business applications, custom software applications, email and messaging. Secondly, there are important findings regarding flow sizes and interarrival times, as well as traffic locality. The number of active flows per second is under 10,000 per rack. 80\% of the flows are smaller than 10KB in size. 80\% of the flows have interarrival times of less than 1\millis in private enterprise \dcs, while 80\% of the flows in university \dcs and in most \dcs have interarrival times between 4\millis-40\millis. 80\% of flows are less than 11s long. Traffic originating from a rack has an ON/OFF pattern with properties that fit heavy-tailed distributions, and traffic that leaves the edge switches is bursty. In cloud \dcs, 80\% of the traffic coming from servers stays within the rack, because administrators colocate dependent applications, whereas in the case of university and private enterprise data centres 40-90\% of the traffic leaves the rack. Link utilisation is higher in the core layer, while the edge layer is lightly utilised. A maximum of 25\% of the core links are highly utilised (hot-spots). Losses are not correlated with high link utilisation, but are due to temporary bursts. Lastly, time of day/week influences link utilisation, especially in the core, and to a moderate degree in the other levels of the \dc.

The second main study about data centre traffic is~\cite{dctraffic}. Over 1 PB of measurement data was collected from 1,500 servers. The workloads were MapReduce style jobs using the Cosmos distributed file system. The servers were instrumented, and then socket level logs were collected along with user application logs. Similarly to the characteristics found in \cite{benson} for cloud \dcs, jobs that require high bandwidth are placed near each other (on the same server, in the same rack, in the same VLAN) by the internal placement algorithm. Regarding traffic locality, the paper observes that there is a probability of 89\% that servers within the same rack do not exchange traffic and 99.5\% in the case of servers in different racks. Regarding congestion, 86\% of the links have congestion periods of at least 10s, while 15\% of the links have congestion periods of at least 100s. Over 90\% of the congestion periods are less than 2s, but more than 1s. Regarding flow sizes, like the previous study~\cite{benson}, most flows were short, 80\% of the flows lasting less than 10s. Less than 0.1\% last longer than 200s. However, more than half of the bytes are found in flows that last less than 25s. Another observation is that there is significant variability in the traffic matrix, both in magnitude and in the pairs of servers which exchange data. Even if the total traffic exchanged remains the same, the pairs of servers involved in this exchange change considerably. Moreover, the traffic experiences periodic short-term bursts, with the interarrival time 15\millis at servers and ToR switches. The median arrival rate of all flows is 100 flows/ms. Lastly, there was no evidence of incast in the cluster.

\begin{table*}
\scriptsize
	\begin{tabular}[c]{|p{1cm} ||p{2.2cm} |p{2.6cm}| p{2.6cm} |p{2.6cm}| p{2.6cm}| }
\hline
\textbf{Study} & \textbf{Concurrent flows} & \textbf{Interarrival times}  & \textbf{Link utilisation} & \textbf{Communication between servers} & \textbf{Other} \\ 
\hline
\hline

\cite{benson}
&
\tabitem the number of active flows per second is under 10,000 per rack \newline

& 
\tabitem 80\% of the flows have interarrival times of less than 1 ms in private enterprise DCs \newline
\tabitem 80\% of the flows in university DCs and in most DCs have interarrival times between 4\millis-40\millis.\newline 
\tabitem traffic originating from a rack has an ON/OFF pattern with properties that fit heavy-tailed distributions \newline 
\tabitem traffic that leaves the edge switches is bursty. 
&
\tabitem link utilisation is higher in the core, while the edge is lightly utilised \newline
\tabitem a maximum of 25\% of the core links are highly utilized (hot-spots) \newline
\tabitem losses are not correlated with link high utilisation, but are due to temporary bursts.
&
\tabitem in cloud data centres, 80\% of the traffic coming from servers stays within the rack \newline
\tabitem in the case of university and private enterprise data centres 40\%-90\% of the traffic leaves the rack. 
&
\tabitem time of day/week influences link utilisation especially in the core and moderate in the other levels of the DC.

\\
\hline
\cite{dctraffic} &
-
&
\tabitem periodic short-term bursts \newline
\tabitem interarrival time 15\millis at servers and ToR switches \newline 
\tabitem the median arrival rate of all flows is 100 flows/ms
&
\tabitem 86\% of the links have congestion periods of at least 10 seconds \newline
\tabitem 15\% of the links have congestion periods of at least 100 seconds \newline
\tabitem over 90\% of the congestion periods are less than 2 seconds and more than 1 second. 
&
\tabitem jobs requiring high bandwidth are placed near each other (on the same server, in the same rack, in the same VLAN) \newline
\tabitem probability of 89\% that servers within the same rack do not exchange traffic and 99.5\% in the case of servers in different racks. 
&

\tabitem significant variability in the traffic matrix, both in magnitude and in the pairs of servers which exchange data \newline
\tabitem no evidence of incast

\\
\hline
\cite{vl2} 
&
-
&
\tabitem a machine has 50\% of the time about 10 concurrent flows \newline
&
-
&
-
& 

\tabitem lack of traffic predictability, no stable traffic matrix

\\
\hline
\cite{alizadeh2010dctcp}
& 
\tabitem median number of concurrent flows per server is 36 \newline
\tabitem 99.99th percentile number of concurrent flows per server is more than 1600
& 
\tabitem the variance in interarrival time is very high, with a very heavy 
tail; spikes occur; large number of outgoing flows happen periodically \newline 
& 
-
&
-
& \tabitem division between query traffic (latency critical) and background traffic \newline

\\
\hline
\cite{augmenting}
&
-
&
-
& 
\tabitem only a few ToR pairs send or receive a large volume of traffic \newline
\tabitem these ToRs exchange much of their data with few of the other ToRs  \newline
\tabitem over 60\% of the demand matrices have fewer than 10\% of their links hot at any time \newline
\tabitem hot links are associated with a high fan-in (or fan-out) \newline
\tabitem fewer than 10\% of hot links repeat \newline 
&
-
&
-

\\
\hline
\cite{infocomvm} & 
\tabitem 80\% of VMs have average traffic rate (over two-week period) less than 800 KBytes/min \newline
\tabitem  4\% of VMs have average traffic rate (over two-week period) 8000KBytes/min \newline
& - 
& - &
-
&

\tabitem overall stable per-VM traffic at large timescales ($> 15$ min) for more than 82\% of the VMs \newline
\tabitem weak correlation between traffic rate and latency   

\\
\hline
\cite{roy2015} &
\tabitem Web servers and cache servers have 100s to 1000s of concurrent connections;  \newline
\tabitem Hadoop nodes have 25 concurrent connections on average;\newline
&
\tabitem median flow interarrival times are 2ms for Hadoop;\newline
\tabitem median flow interarrival times are 3ms and
8ms for cache leaders and followers;
&
\tabitem between hosts and the ToR is quite low;\newline
\tabitem the median utilisation between the ToR and aggregation switches is between 10-20\% across clusters;\newline 
\tabitem between the aggregation and core switches the utilisation is even higher;\newline
\tabitem load varies significantly between clusters; \newline
&
\tabitem cache followers and leaders communicate
with 175-350 different racks concurrently;\newline
\tabitem Web servers communicate with 10-125 racks concurrently; \newline
\tabitem most of the traffic is destined to only a few 10s of racks; \newline
&
\tabitem load balancing is effective;\newline
\tabitem traffic demands are stable over intervals as long as 10 seconds;\newline
\tabitem heavy hitters' sizes are not much larger than the median flow sizes, and they change rapidly;\newline
\tabitem packet sizes are small, median length for non-Hadoop traffic $<$ 200 bytes.
\\
\hline
\end{tabular}
\caption{Data centre network traffic characteristics - part 2.\label{tab:dc-traffic-characteristics-2}}
\end{table*}

The most recent study is on Facebook's \dc network traffic characteristics~\cite{roy2015}. The data was collected using \emph{Fbflow} (an internal monitoring system that samples packet headers with a sampling rate of 1:$30,000$), through port mirroring on the ToR, and mirroring of all traffic from one server. The traffic is one of the following: Web, MapReduce, MySQL, or traffic served from cache servers (leader and follower). 
Network traffic characteristics regarding flow sizes and traffic locality for Hadoop jobs are similar to the ones found by the previously mentioned studies. However, the traffic patterns for the other types of applications differ substantially from the ones described in~\cite{dctraffic,benson}. The majority of traffic is intra-cluster (57.5\%, from caching follower servers), with only 12.9\% intra-rack. Also, there is a significant portion of intra-datacentre and inter-datacentre traffic, in particular from the caching leader servers. The Hadoop traffic is more rack-local than other types of applications. Frontend traffic has minimal rack-local traffic, but significant intra-cluster traffic. Overall, the locality patterns are stable over time periods ranging from seconds to days. Regarding flow sizes, most Hadoop flows are short, while for the other types of services, they are long-lived, but internally bursty and do not carry a significant number of bytes. Cache flows are larger than Hadoop flows, and Web server flows are between the two. Facebook's use of load balancing is effective. It distributes the traffic across hosts, except in the case of Hadoop servers, which see jobs of different sizes, and traffic demands are quite stable over sub-second intervals. 
Consequently, heavy hitters' sizes are not much larger than the median flow sizes, and they change rapidly, making it hard to predict them.
Packet sizes are small, median length for non-Hadoop traffic being less than 200 bytes, while for Hadoop the distribution is bimodal (1500 bytes or TCP ACKs size). The traffic does not exhibit ON/OFF arrival behaviour, unlike the traffic in the previous studies~\cite{dctraffic, benson}. Web servers and
cache servers have 100s to 1,000s of concurrent connections, while Hadoop nodes have
25 concurrent connections on average, similar to the values reported in~\cite{dctraffic}.
Median flow interarrival times are 2\millis for Hadoop, and 3\millis and 8\millis for cache leaders and followers respectively.
Cache followers and leaders communicate with 175-350 different racks concurrently, while Web servers communicate with 10-125 racks. However, most of the traffic is destined to only a few 10s of racks. Link utilisation on links between hosts and the ToR is quite low, with the average 1-minute link utilisation
less than 1\%. Load varies significantly between clusters, a Hadoop cluster being five times more loaded than a Frontend cluster. The median utilisation between the ToR and aggregation switches is between 10-20\% across clusters. At this level, the
difference between clusters is not as significant as in the previous case, with the most loaded clusters being three times more loaded than the lightly loaded ones. Between the aggregation and core switches the utilisation is even higher.

\textbf{\emph{Lessons Learned:}} To sum up, \dc traffic characteristics depend on the type of applications deployed, \eg, Hadoop, Web, caches, on diurnal patterns, and on the \dc operators' strategies for application placement and load balancing.

\section{Cluster workloads in \dcs}
\label{sec:cluster-workloads}

We start by defining common terms used in the cluster scheduling literature.
An application is called a \emph{job}. A job may have multiple \emph{tasks}. A task is an application instance of the job represented by one or multiple processes that run inside a container or virtual machine, usually on a single core. 
The tasks of a job must be placed on the available machines. A~\emph{cluster scheduler} decides on which machines to place the tasks 
of the jobs. Cluster scheduling in its simplest form is bin-packing of tasks on the available machines.
However, this simple allocation mechanism might not yield the optimal performance for applications due to lack of adequate resources, or possible interference between tasks that share the underlying host hardware and network. These issues can be solved by respecting the job's resource demands or by defining placement constraints. Resource demands usually comprise the number of cores, the amount of memory, disk throughput, or network bandwidth that a task needs. Placement constraints can be defined to avoid colocation between tasks that might interfere, or to allocate a task to a machine with certain characteristics. 
Solving these issues has given rise to a large body of work on how to best map job demands and constraints to job allocation systems. 

We next discuss the characteristics of the main cluster workloads released by companies in the past years~\cite{google-workload, cortez2017, cluster-workloads}. 

\textbf{Google workload} The most well-known cluster trace is from Google~\cite{google-workload}. It is a 2011 cluster trace from a $12,500$ machines cluster. The Google workload is a 29-day trace of jobs that run on bare-metal hosts.
Task runtimes are not uniform, with $80$\% of tasks running for less than 12 minutes~\cite{cluster-workloads}. A similar observation can be done with regards to task resource requests, where $90$\% of the smallest jobs request 16 CPU cores or fewer~\cite{cluster-workloads}. The trace has sub-second job interarrival times. An updated trace from 2019 was released~\cite{borg-next}. The authors compare the two traces, finding that the workload arrival rate has increased, with the job submission rate $3.7$ times higher than in 2011. The workload mix has also changed, with jobs having moved from the free tier to the best-effort batch tier. 
The workload continues to exhibit a heavy-tailed distribution where the top 1\% use up over 99\% of resources. 
\textbf{Microsoft Azure workload} The Microsoft Azure VM workload~\cite{cortez2017} is the first of its kind publicly released. It spans three months, and it includes first-party workloads (internal VMs and first-party services offered to third-party customers), and third-party workloads (VMs created by external customers). More than $90$\% of VMs run for less than a day, and a small percentage of long-running VMs use up more than $95$\% of the total core hours. 
In terms of VM core count, almost $80$\% of VMs have a maximum of two cores, with almost $60$\% of VMs using only one core. In terms of memory, $70$\% of VMs use less than $4$ GBytes. In terms of deployment sizes, around $40$\% have a single VM and $80$\% have at most $5$ VMs. Regarding VM workloads, $68$\% of core hours are categorised as delay-insensitive (batch workloads, internal workloads), and around $28$\% are interactive, while the remaining $4$\% are not categorised. The VM arrival times are bursty and diurnal, and there are less VMs running during the weekend.

\begin{figure}
 \centering
        \includegraphics[width=0.5\textwidth]{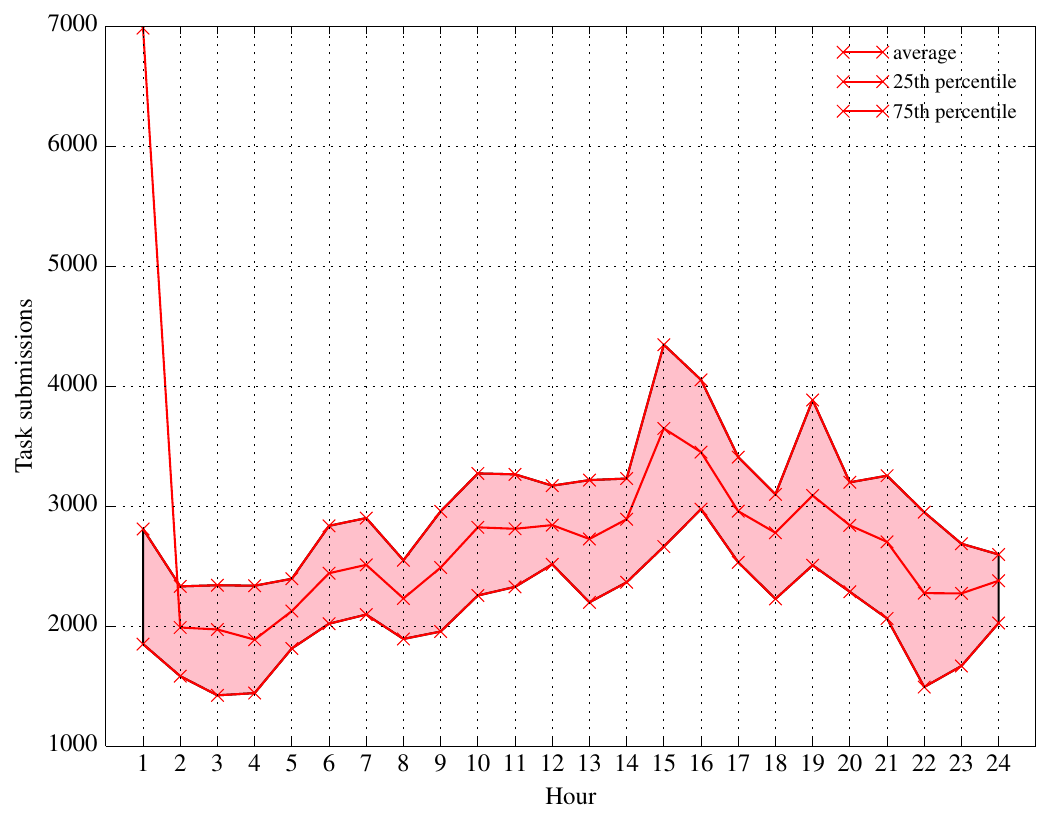}
        \caption{Azure workload number of task arrivals per hour - average, 25\thss and 75\thss percentiles.}
       \label{fig:azure-task-arrivals-per-hour}
\end{figure}

Both Google and Azure workloads have a high number of tasks placed (more than 140K tasks) at the beginning of the trace (timestamp 0) to setup the cluster state. Afterwards, the average number of task arrivals per hour is between 1800-3500 for the Azure trace (Figure~\ref{fig:azure-task-arrivals-per-hour}), and 40K-70K for the Google trace~\cite{cluster-workloads}.

\textbf{Two Sigma and Los Alamos National Laboratory workloads} These traces share some of the characteristics of the Google workload. They have sub-second job interarrival times, requiring sub-second scheduling decisions.
On the other hand, they display diurnal patterns in job submissions, similar to the Microsoft Azure workload, but not present in the Google workload.
While most of the jobs have short durations and request a small number of cores in the Google workload, this is not true for the Two Sigma and LANL traces. The median Google job is $4-5\times$ times shorter, and requests $3-406\times$ fewer CPU cores.
Similarly, most of the VM deployments in the Microsoft Azure workload are small, with $80$\% of VMs having a maximum of two cores. 
Regarding job duration, $80$\% of Google jobs are less than 12 minutes, whereas in the other traces the same fraction of jobs are several hours long (2-6 hours). The Google workload has a long tail, with some jobs running for at least the duration of the whole trace. 

\textbf{\emph{Lessons Learned:}} A commonality of the cluster workloads released is that they do not include information related to networking demands, \eg, network bandwidth, or how latency-sensitive the application is. The closest that is available is the Microsoft Azure workload, which mentions which VM workload is interactive. Still, this information is insufficient, since it does not offer a quantifiable metric, \eg, what is the job's SLO.
Therefore, this lack of information represents a challenge when developing new cluster scheduling policies that try to improve application performance while considering networking demands. 

\section{Network monitoring systems}
\label{sec:network-monitoring}

Table~\ref{tab:network-monitoring} summarizes the network monitoring systems presented in this section.

\subsection{Network monitoring in SDN-enabled networks}
\label{label:SDN}
Software Defined Networking (SDN) is a paradigm in which the control plane is separated from the data forwarding plane, enabling the centralisation of network control and offering the possibility of programming the network. The control plane is represented by a \emph{controller} and the data plane consists of networking devices, such as switches and routers. This separation is made possible by a programming interface, which allows the controller to communicate with the forwarding devices, for example to install forwarding rules at switches. OpenFlow~\cite{OpenFlow} is the most popular such application programming interface (API). In an OpenFlow network, the controller can collect statistics about the flows (duration, number of packets, number of bytes) by polling the switches. These statistics can be either per-flow values or aggregates across multiple flows that match a rule. SDN monitoring tools use flow events that the controller receives 
and the statistics collected by the controller. We discuss several such tools in the following paragraphs.

OpenTM \cite{OpenTM} is a traffic matrix (TM) estimator for OpenFlow networks. It determines the current active flows in the network based on flow initiation and termination events. OpenTM uses routing information from the OpenFlow controller to discover the flow paths, and then it periodically polls switches on the flow path, obtaining byte and packet counters. OpenTM assumes that all the packets of a flow follow the same path in the network. 

FlowSense~\cite{FlowSense} is a monitoring tool that uses OpenFlow control messages to determine average link utilisation in OpenFlow-enabled networks. When a flow expires, the controller receives information about the duration and size of that flow. The link utilisation is computed only at certain times, \eg, when all the flows on a link have expired. In the case of long flows, or if rules have large timeouts, the link utilisation is rarely computed. While the FlowSense approach does not present any overhead in terms of additional messages injected in the network,
the delay between obtaining average link utilisation estimates can be 10 seconds with 90\% accuracy. 
Given these limitations, an adaptive monitoring method was proposed in PayLess~\cite{PayLess},
which additionally polls the switches. 
In doing so, it obtains flow statistics more often than FlowSense, which gathered statistics only when a flow terminated. 

DREAM~\cite{dream} is a network-wide measurement architecture that uses OpenFlow to coordinate the measurement devices. DREAM implements an algorithm for allocating switch memory 
resources depending on the measurement task needs, traffic and expected accuracy, multiplexing the resources both temporally and spatially. 

OpenSketch~\cite{opensketch} is a measurement architecture inspired by the software-defined networking paradigm, dubbed software-defined traffic measurement. In the switch data plane, a packet goes through $3$ basic blocks: hashing (certain packet fields are hashed), classification (matching on fields
according to predefined rules), 
and counting (gathering statistics).
The control plane manages the data plane, and configures the measurement tasks according to available resources. 
OpenSketch uses sketches, data structures from streaming algorithms, to store information about packets. 
Sketches have two main advantages compared to flow-based counters: low memory usage and the possibility of setting the desired accuracy in relation to the memory used. 

SLAM~\cite{slam} is a latency monitoring framework for SDN-enabled \dcs. It sends probe packets to trigger control messages from the first and the last switch of a network path. Based on the arrival times at the controller of the control messages, it computes a latency distribution for that network path, being able to detect increases in latency of tens of milliseconds. 

\textbf{\emph{Lessons Learned:}} The information and the granularity provided by OpenFlow counters is limited, hence other approaches sought to use known ways of acquiring statistics from the network (\eg, sampling~\cite{OpenSample}, port mirroring~\cite{Planck}). Still, SDN monitoring has some benefits. It offers the possibility of greater control and reduced human overhead for configuring each switch individually to collect network statistics. Another benefit of SDN for monitoring is the centralised view of the network, which allows for a better allocation of network resources~\cite{dream} to gather data. Also, the centralised view can help in reducing the collection of redundant data (\eg, the same flow being sampled at several places in the network). These aspects are exploited in some network monitoring frameworks for data centres. 

\begin{table*}[!t]
\begin{tabular}{|p{65mm}||c|p{65mm}|}
\hline
\textbf{Study} & \textbf{Measurement techniques} & \textbf{Measurement task} \\
\hline
\hline
OpenTM~\cite{OpenTM} & OpenFlow & traffic matrix \\
\hline
FlowSense~\cite{FlowSense}, PayLess~\cite{PayLess} & OpenFlow & link utilization \\
\hline
SLAM~\cite{slam} & OpenFlow & path latency \\
\hline
DREAM~\cite{dream} & OpenFlow & heavy hitter detection, hierarchical heavy hitter detection, anomaly detection \\
\hline
OpenSketch~\cite{opensketch} & FPGA & heavy hitters, superspreader/DDoS, traffic changes detection, flow size distribution, count traffic \\
\hline
NetNORAD~\cite{netnorad} & UDP probes & RTT; packet loss \\
\hline
Everflow~\cite{everflow} & TCP probes & RTT \\
\hline
Pingmesh~\cite{pingmesh} & TCP probes & RTT; packet drops\\
\hline
VNET Pingmesh~\cite{vnet-pingmesh} & TCP probes & RTT \\
\hline
PTPmesh~\cite{diana-mascots, diana-tma, diana-tnsm} & UDP probes & one-way delay; packet loss\\
\hline
UnivMon~\cite{univmon}, ElasticSketch~\cite{elasticsketch}, SketchLearn~\cite{sketchlearn}, NitroSketch~\cite{nitro}, CocoSketch~\cite{coco}, HashPipe~\cite{hashpipe}, ElasticTrie~\cite{diana-sosr, diana-sosr2}, FlowRadar~\cite{flowradar}  & Programmable switches & heavy hitters, change detection, DDoS, flow size distribution, flow size estimation \\
\hline
Marple~\cite{marple}, LossRadar~\cite{lossradar} & Programmable switches & packet loss; packet latency per flow \\
\hline 
In-band Network Telemetry (INT)~\cite{int}, PINT~\cite{pint} & Programmable switches & path latency \\
\hline
\end{tabular}
\caption{Network monitoring systems.}
\label{tab:network-monitoring}
\end{table*}

\subsection{Data centre network monitoring systems}
\label{sec:dc-measurement-tools}

\textbf{A \dc network monitoring system} needs to not use many resources (both at the servers and in-network), should be always-on and it should be able to monitor the whole network in a scalable manner.
Large-scale \dc monitoring systems, such as NetNORAD~\cite{netnorad}, Everflow~\cite{everflow}, Pingmesh~\cite{pingmesh}, VNET Pingmesh~\cite{vnet-pingmesh},
have been deployed by cloud providers in their \dcs.
NetNORAD~\cite{netnorad} is a system used in Facebook's \dcs to measure RTT and packet loss ratio, and uses ping between servers. 
Everflow~\cite{everflow} is a system that monitors control packets and special TCP packets (TCP SYN, FIN, RST). Additionally, it supports guided probing by injecting custom packets, which are used to measure link RTTs.
Pingmesh~\cite{pingmesh} is a system that runs RTT measurements between every two servers in \dcs, and it measures inter-server latencies at three levels: Top-of-Rack switch, intra \dc and inter \dc. Also, the system reports the packet drop rate, which is inferred based on the TCP connection setup time.
VNET Pingmesh~\cite{vnet-pingmesh} monitors latency for tenant virtual networks (VNETs),
with the TCP probes being sent from the virtual switch at the end-host. 
PTPmesh~\cite{diana-mascots, diana-tma, diana-tnsm} is a cloud monitoring tool that uses the Precision Time Protocol (PTP) statistics to infer the one-way delay and packet loss in \dcs.

\textbf{Fault localisation} A common goal of most of these large scale measurement systems is fault localisation. For example, NetNORAD is used in conjuction with fbtracert~\cite{netnorad}, which traces multiple paths between two endpoints in the network in parallel to determine the location of the fault. Pingmesh is used to detect switch silent packet drops. 
NetPoirot~\cite{arzani2016} presents a classification algorithm that identifies the root cause of failures using TCP statistics collected at one of the endpoints. 
The work in~\cite{arjun2017} looks from the end-host to identify the faulty links and switches, by correlating anomalies in end-host statistics with the network path of the packets. 
007~\cite{007} tracks the path of TCP connections that display retransmissions through traceroute, and identifies the links with the most retransmissions as the faulty ones. PathDump~\cite{pathdump} traces packets through \dc networks, and can report poor TCP performance when the number of consecutive
packet retransmissions is above a certain threshold for a flow.~\cite{zhang2017} presents a sampling framework which can poll a subset of switch counters at microsecond-level granularity to determine microbursts in \dcs.

\textbf{Programmable data planes} The monitoring approaches that use SDN presented in~\ref{label:SDN} do not have the granularity necessary for data centres. This drawback lead to the increasing use of programmable switches~\cite{bosshart} for monitoring tasks~\cite{flowradar, lossradar, univmon, int, diana-sosr, diana-sosr2, hashpipe, elasticsketch, sketchlearn, sonata, nitro, coco}. 
FlowRadar~\cite{flowradar} keeps track of all the flows
in the network with their associated counters, and exports this
information periodically to a remote collector, which ultimately
uses them for different monitoring applications targeted to datacenters. 
UnivMon~\cite{univmon}, ElasticSketch~\cite{elasticsketch}, SketchLearn~\cite{sketchlearn}, NitroSketch~\cite{nitro}
and CocoSketch~\cite{coco} use sketch-based data structures in
the dataplane to record network traffic statistics that are exported
at fixed time intervals to the control plane that processes them to perform different measurement tasks.
HashPipe~\cite{hashpipe} and ElasticTrie~\cite{diana-sosr, diana-sosr2} focus on determining the largest flows (heavy hitters).
Sonata~\cite{sonata} proposes a query interface for network telemetry,
uses sketches in the dataplane, and the controller zooms-in the network
traffic of interest by refining the network query.
In-band Network Telemetry (INT)~\cite{int} measures the end-to-end latency between virtual switches, by appending the switch per-hop latency to the packet that traverses the network path between the two virtual switches. The end-to-end latency is computed by adding the per-hop latencies, while the propagation delays are considered negligible. 
PINT~\cite{pint} is also an in-band network telemetry framework, but it bounds the amount of measurement information added to each packet by enconding it across multiple packets. PINT can compute tail latencies and can help applications such as path tracing and congestion control.  
LossRadar~\cite{lossradar} detects packet losses in \dcs within tens of milliseconds, and it reports the 5-tuple flow identifiers and the switches where these losses were detected.
Marple~\cite{marple} uses a programmable key-value store implementation on switches for measurements such as packet loss per connection, packet latencies per flow, or determining packets that have high end-to end-queueing latency. 

\textbf{\emph{Lessons Learned:}} Programmable switches offer high-speed, complex monitoring and fault localisation functions. The network monitoring information can be leveraged further for resource allocation and control. One drawback is that legacy switches have to be replaced with programmable switches to be able to take advantage of this rich set of monitoring functions, and this can be costly in the short-term for companies. On the other hand, simple network monitoring techniques such as active probing and end-host and legacy switches statistics coupled with SDN can still provide benefits to \dc operators.
\section{Cluster scheduling}
\label{sec:bk-cluster-scheduler}

Most of the cluster schedulers take into account any other constraints but those related to networking requirements or the state of the network. 
These constraints generally represent the needs of a job with respect to the number of cores, CPU utilisation, memory, affinity with other tasks, data locality, placement constrains, and low placement latency. We call these cluster schedulers \emph{conventional}. \emph{Network-aware} cluster schedulers try to ensure tail latency and bandwidth guarantees, incorporating such constraints in their placement algorithms. Recently, \emph{ML-driven} cluster schedulers have been proposed, however their current algorithms do not consider networking monitoring data or resources as an input. We discuss these categories in the next sections.

\subsection{Conventional cluster schedulers}
\label{sec:conventional}

In Section~\ref{sec:conventional}, we discuss the features of the most important conventional cluster schedulers. 
Gog~\cite{gog} and Schwarzkopf~\cite{malte} identify the most important features of cluster schedulers: \emph{architecture}, \emph{multi-dimensional resource allocation}, \emph{resource allocation and dynamic adjustment}, \emph{task interference}, \emph{constraint handling}, \emph{data locality} and \emph{low placement latency}.

\textbf{Scheduler architecture} There are several types of cluster scheduler architectures.

\emph{Centralised schedulers}, such as Borg~\cite{borg}, Bistro~\cite{bistro}, Quincy~\cite{quincy}, Firmament~\cite{firmament}, TetriSched~\cite{tetrisched}, have the advantage of having access to all of the information related to the cluster's state (\eg,~where the tasks are running, which tasks should be scheduled, and which machines have free cores). Consequently, these schedulers compute high-quality task placements. On the other hand, task placement latency can be significant because the computation delay grows in proportion to the cluster size~\cite{quincy}. Significant task placement latency can be detrimental for short-running tasks that might end up waiting to be scheduled for a time greater than their actual duration. 

\emph{Distributed schedulers} (Apollo~\cite{apollo}, Sparrow~\cite{sparrow}) can solve this issue, but since they do not have the full view of the cluster state, they trade off high-quality task placements for lower task placement latency. 

\emph{Hybrid schedulers}' architecture is comprised of a centralised scheduler and one or more distributed schedulers. Examples of hybrid schedulers are Hawk~\cite{hawk}, Eagle~\cite{eagle}, Mercury~\cite{mercury}. The long-running tasks are usually placed by the centralised scheduler, while the short tasks are placed by the distributed schedulers. 

Another architecture is that of \emph{two-level schedulers} (Mesos~\cite{mesos}, YARN~\cite{yarn}), which have an application-level scheduler that maps the tasks to the resource allocations determined by a resource scheduler. For example, the MapReduce jobs are scheduled by YARN's job manager, while resources are allocated by YARN's resource manager.

\textbf{Multi-dimensional resource allocation} Nowadays complex \dc applications have different resource requirements (\cref{ss:dc-apps}). Some cluster schedulers~\cite{quincy, jockey, sparrow, hawk, mercury, eagle} allocate the same amount of resources to all tasks regardless of their needs, and this may hurt application performance. Other schedulers~\cite{mesos, yarn, omega, quasar, bistro, apollo, tetris, tetrisched, borg} support specifying job resource requirements. 

\textbf{Resource allocation and dynamic adjustment} In most cases, users can specify their jobs' resource requirements~\cite{borg}, but often overestimate their requirements~\cite{gog}, or they do not know how much to request~\cite{mogul:2015}. 
This can be solved in several ways: i) profiling the job before the actual execution~\cite{quasar}, ii) decreasing or redistributing the initial resource allocation after an observation period~\cite{borg, altruistic}, or iii) analysing historical traces of jobs to build a performance model of the expected performance depending on the job resource allocation~\cite{jockey, morpheus, ernest, cherrypick, optimus}. Some performance models were developed for specific types of jobs (MapReduce and Dryad in Jockey~\cite{jockey}, deep learning jobs in Optimus~\cite{optimus}). 
Also, techniques such as Bayesian optimisation~\cite{cherrypick} or non-negative least square~\cite{ernest, optimus} have been used to build performance models for different applications. 

\textbf{Task interference} To achieve high cluster utilisation, multiple tasks are colocated on each machine. But tasks compete for the same resources (memory, cache, disk, network), and they may then interfere with each other, causing a decrease in application performance. As a consequence, some schedulers (Paragon~\cite{paragon}, Quasar~\cite{quasar}) first determine whether the tasks waiting to be scheduled interfere with each other, and only then place the tasks based on the result. 

\textbf{Constraint handling} Tasks can have different placement constraints. For example, an application's performance would benefit from running on a certain type of hardware, \eg, machine learning tasks require specialist hardware such as General Purpose Graphics Processing Units (GPGPUs) or Tensor Processing Units~\cite{tpu} (TPUs). Constraints can be \emph{hard} (mandatory), \emph{soft} (not mandatory) or \emph{complex} (combination of soft and hard constraints). 

Hard constraints must be satisfied, the tasks not being scheduled until machines that satisfy these constraints are available. Hard constraints usually refer to hardware architecture and kernel version~\cite{sharma:2011}. In the Google cluster workload, approximately 6\% of all tasks have hard constraints. 

Soft constraints, on the other hand, are not mandatory, and tasks can be scheduled to run even if their constraints are not met. Quincy~\cite{quincy} and Firmament~\cite{firmament} model the cluster scheduling problem as a min-cost max-flow optimisation over a flow network. The nodes in the flow network represent the tasks that are submitted and the machines of a cluster. The task and machine nodes are connected through task placement preference arcs which represent soft constraints. A min-cost max-flow algorithm run over the flow network computes task placements.
In Medea~\cite{medea} the constraints are soft by default, and weights can be assigned to the constraints to rate their importance. In Kubernets~\cite{kubernetes}, one can specify if a constraint is hard or soft.

Complex constraints are a combination of hard and soft constraints, and involve satisfying the requirements of multiple tasks or machines.
They are supported by few schedulers, an example of such scheduler being TetriSched~\cite{tetrisched}. 
An important type of complex constraint is task affinity or anti-affinity. Task affinity refers to placing two or more tasks that have a dependency on the same resource. In contrast, task anti-affinity means placing the tasks on different resources. Kubernetes~\cite{kubernetes} supports affinity/anti-affinity constraints. 

Satisfying constraints generally increases task placement latency~\cite{sharma:2011} and limits the scheduler's scalability.

\textbf{Data locality} Data locality used to be an important feature of cluster schedulers~\cite{quincy, corral}, since disk throughput is higher than network bandwidth, it is desirable for the data to be stored as close as possible to the application that uses it.
However, the current \dc networks can provide one Pbps of bisection bandwidth~\cite{singh2015jupiter} and the recent trend in resource disaggregation in \dcs~\cite{gao2016,legoos} 
make this feature less important, since the data is now accesible over a very fast network. Nevertheless, generating less network traffic by reading from local disks in a traditional server architecture lowers network utilisation, and thus reduces network congestion, providing more predictable application performance~\cite{qjump}.

\textbf{Low placement latency} The time it takes to compute a placement for a task is an important feature for a cluster scheduler. A low placement latency 
favours high cluster utilisation, and reduces the waiting time of the tasks before being scheduled. Cluster schedulers that support complex constraints may be slow in placing tasks, but they compute high-quality task placements. Quincy's algorithm~\cite{quincy}, for example, can take minutes for the Google workload on a cluster with 12,500 machines~\cite{firmament}. This runtime is too large for workloads that have sub-second job interarrival times (\cref{sec:cluster-workloads}), and it can lead to an increase in task wait time, which is especially detrimental for short-running tasks. On the other hand, Firmament's algorithm runtime~\cite{firmament} is sub-second for the same scenario, achieving both high-quality placements and low placement latency. Other cluster schedulers use less complex algorithms which take less time, but do not offer high-quality task-placements, since they do not take into account the tasks' constraints. Such an example is Sparrow~\cite{sparrow}, which places short-running tasks in a random manner. 

\subsection{Network-aware cluster schedulers}
\label{sec:network-aware}

Few cluster schedulers consider the networking demands of the applications and the network conditions in the \dc. Managing the network traffic to achieve short flow completion times is left to \dc transport designs~\cite{alizadeh2010dctcp, pfabric, phost, qjump, ndp} and flow schedulers~\cite{fastpass}, this happening after the applications have been scheduled to run in the \dc.
In general, incorporating network demands within the cluster scheduler has been treated as a separate problem from cluster schedulers that take into account only the host resources required by a job. 
Table~\ref{tab:comparison-network-guarantees} describes the mechanisms to allocate network bandwidth between tenants and to provide tail latency guarantees to ensure predictable performance.

\subsubsection{Network bandwidth guarantees} 

In the past, network throughput variability in cloud providers was an important issue, with bandwidth varying by a factor of five in some cases~\cite{ballani2011}, leading to uncertain  application performance and, consequently, tenant cost. The throughput variability was the result of different factors, such as network load, tenant VM placement, and oversubscribed \dc networks. 

Nowadays, \dc networks often utilise full bisection bandwidth~\cite{singh2015jupiter, cortez2017}. Also, cloud providers' commercial offerings list the expected network bandwidth for each type of VM, be it an exact value (Microsoft Azure, Google Cloud Platform) or qualitative estimate (Amazon EC2). These changes fit with the observation that network bandwidth guarantees have improved in recent years. Recent studies have shown that cloud providers like Amazon EC2~\cite{persico-ec2} and Microsoft Azure~\cite{persico-azure} see less throughput variability. In the case of Microsoft Azure, larger VMs and which are placed within the same affinity group or virtual network have better network throughput and observe small variability, whereas medium sized VMs experience higher variability regardless of the policy applied, with some regions offering better performance than others. On the other hand, $60$\% of the tenants of Microsoft Azure use the smallest VM size (1 core) and $20$\% the medium VM size (2 cores)~\cite{cortez2017}, which means that most of the tenants do not have strict network bandwidth guarantees even in today's \dcs. In the case of Amazon EC2~\cite{persico-ec2}, the network throughput is stable over time regardless of VM size, larger VMs can achieve higher throughput compared to smaller ones, and there was no difference between regions observed.

Allocating network bandwidth between endpoints was first described in the context of Virtual Private Networks (VPN)~\cite{hose}. In the cloud computing model, the VPN customers can be assimilated with the tenants from the cloud, and a VPN endpoint's equivalent is a VM.
The customer-pipe model is the allocation of bandwidth on paths between source-destination pairs of endpoints of the VPN. In this model, a full mesh between customers is required to satisfy the SLAs. In the hose model, an endpoint is connected with a set of endpoints, but the bandwidth allocation is not specified between pairs. Instead, the aggregate bandwidth required for the outgoing traffic to the other endpoints and the aggregate bandwidth required for the incoming traffic from the other endpoints in the hose is specified.
These two models served as basis for bandwidth allocation in the cloud, with the hose model being frequently used~\cite{secondnet, gatekeeper, ballani2011, eyeq}. SecondNet~\cite{secondnet} introduces an abstraction called virtual data centre (VDC) and multiple types of services (type 0 guaranteed bandwidth between VMs, type 1 per ingress/egress bandwidth reservation for VM, and best effort).
Gatekeeper~\cite{gatekeeper} supports the hose model and sets minimum bandwidth guarantees for sending and receving traffic for a VM, which can be increased up to a maximum rate if unused capacity is available.
It also proposes an extension to the hose model by composing multiple hoses for a VM to incorporate application communication patterns. EyeQ~\cite{eyeq} uses a similar mechanism to Gatekeeper.

Several works extend the hose model.  
ElasticSwitch~\cite{elastic} provides minimum bandwidth guarantees by dividing the hose model guarantees into VM-to-VM guarantees, and taking the minimum between the guarantees of the two VMs. It rate limits the traffic from a VM to the specified guarantee or higher if there is available capacity. The unused capacity on a link is allocated proportionally to the bandwidth guarantees of the VM pairs using that link.
Oktopus~\cite{ballani2011} uses the hose model (named virtual cluster, where all the VMs are connected through a single switch) and virtual oversubscribed cluster model (groups of virtual clusters connected through a switch with an oversubscription factor).
Proteus~\cite{proteus} authors analyse the traffic patterns of several MapReduce jobs, finding that there is no need to allocate a fixed network bandwidth from the start to the end of the job, because the network demands of the applications change over time. They propose a time-varying network bandwidth allocation scheme: temporally interleaved virtual cluster, which is a variant of the hose model.
CloudMirror~\cite{cloudmirror} derives a network abstraction model, tenant application graph, based on the application's communication pattern. The applications considered in this work usually have multiple tiers or components, and each tier/component is formed of a number of VMs. Bandwidth within the component is allocated using the hose model. Bandwidth between components is allocated by guaranteeing each VM in a component $C_{1}$ a send bandwidth to send traffic to the VMs in a component $C_{2}$, and each VM in component $C_{2}$ is guaranteed a receive bandwidth to receive traffic from VMs in component $C_{1}$. Pulsar~\cite{pulsar} provides end-to-end isolation for VMs and appliances (\eg,~load balancing, storage, monitoring). It forms a virtual \dc out of dedicated appliances connected to VMs through virtual switches, where each link between VMs and appliances has a throughput guarantee. 
Implementation-wise, most of the works enforce rate limits at the end-host's hypervisor.

\begin{table*}
\scriptsize
\centering
\begin{tabular}{|p{13mm}||p{20mm}|p{12mm}|p{13mm}|p{13mm}|p{16mm}|p{18mm}|p{10mm}|p{6mm}|}
\hline
& \textbf{BW Guarantees} & \textbf{Work-conserving} & \textbf{Topology} & \textbf{Adaptability} & \textbf{Communication Pattern} & \textbf{Implementation} & \textbf{Latency Guarantees} & \textbf{VM placement} \\
\hline
\hline
SecondNet \cite{secondnet} & Yes, Hose, VM-to-VM & No & No & No & No & Hypervisor, Source routing, MPLS, Central Controller & No & Yes \\
\hline
Gatekeeper \cite{gatekeeper}, EyeQ \cite{eyeq} & Yes, Hose & Yes & No congestion in the core & Yes & Yes & Hypervisor & No & No\\ 
\hline
Seawell \cite{seawell} & No & Yes & No & No & No & Hypervisor & No & No\\
\hline
Oktopus \cite{ballani2011} & Yes, Hose, Virtual oversubscribed cluster & No & Tree & No & No & Hypervisor, Central Controller & No & Yes \\ 
\hline 
Proteus \cite{proteus} & Yes, Temporally interleaved virtual cluster & No & Tree & Yes & Yes & Profiling, Central controller & No & Yes\\
\hline 
NetShare \cite{netshare} & No & Yes & No & No & No & Central Controller, Switches & No& No\\
\hline
PS-P \cite{faircloud} & Yes, Hose, Virtual oversubscribed cluster & Yes & Tree, Fat-tree & No & Number of VMs that communicate with a VM &  Switches &  No & No\\
\hline
ElasticSwitch \cite{elastic} & Yes, Hose & Yes & No & Yes & Yes & Hypervisor & No & No\\
\hline
Hadrian \cite{ballani2013} & Yes, Hose & Yes & Tree & No & Yes & Hypervisor, switches, central controller & No & Yes\\
\hline
Choreo \cite{choreo} & Yes, Hose, VM-to-VM & No & Yes & No & Yes &  Profiling, measurement, placement components & No & Yes \\
\hline
CloudMirror \cite{cloudmirror}  & Yes, Tenant Application Graph & No & Tree & No & Yes & Central controller, ElasticSwitch & No & Yes\\
\hline
Pulsar \cite{pulsar}  & Yes, Virtual \dc & Yes & No & Yes & No & Central controller, Hypervisor & No & No\\
\hline 
Cicada \cite{cicada}  & No & No &  No & Yes & Yes & Hypervisor, Switches & No & Yes \\
\hline
Silo \cite{silo} & Yes & No & Tree & No & No & Central controller, End-host rate limiter & Yes, tail & Yes \\
\hline
QJump \cite{qjump} & Yes & No & No & No & No & End-host rate limiter & Yes, tail & No \\
\hline
SNC-Meister \cite{sncmeister} & Yes & No & No & No & Yes, traces & Centrall controller and end-host rate limiter & Yes, tail & Yes \\
\hline
NoMora \cite{diana-nomora} & - & - & - & Yes & Yes & Cluster scheduler & Yes & Yes \\
\hline
\end{tabular}
\caption{Systems providing network bandwidth and tail latency guarantees in \dcs.}
\label{tab:comparison-network-guarantees}
\end{table*}
 
Profiling applications to determine their network throughput and keeping historical network throughput values are two aspects that can help to allocate bandwidth in a more efficient manner~\cite{proteus, cicada}. 
Using real-time measurements of the network conditions, in particular throughput, has been employed to improve VM placement~\cite{choreo}. 
Choreo~\cite{choreo} measures the network throughput between VM pairs through packet trains, estimates the cross traffic, and locates bottleneck links. Based on the network measurements and application profiles (number of bytes sent), it makes VM placement decisions to minimise application completion time.
The NoMora cluster scheduling architecture~\cite{diana-nomora} uses network latency measurements between pairs of hosts and application performance predictions dependent upon current network latency to decide where to place tenants' VMs. The authors determine experimentally the relationship between network latency and application performance for typical cloud applications. This is similar to the profiling phase of applications and building a performance model done by different cluster schedulers, but these frameworks did not consider network latency demands in their profiling phase, nor in their performance model.

VM placement to provide network bandwidth guarantees starts by looking at subtrees in the topology to place the VMs, and goes upward in the tree to find a suitable allocation~\cite{ballani2011, proteus, ballani2013, cloudmirror}. This naturally leads to the VMs being allocated within the same rack or within the same pod, which may hurt application availability if the links to the servers that run the application fail~\cite{cloudmirror}. To mitigate this, CloudMirror~\cite{cloudmirror} additionally incorporates an anti-affinity (anti-colocation) constraint in the VM placement algorithm. SecondNet~\cite{secondnet} builds a bipartite graph whose nodes are the VMs on the left side and the physical machines on the right side, and then finds a matching based on the weights of the edges of the graph using min-cost max-flow. The weights are assigned based on the available bandwidth of the corresponding server. Hadrian~\cite{ballani2013} provides bandwidth guarantees for inter-tenant communication. The VM placement algorithm builds a flow network to express the VMs communication patterns and minimum bandwidth constraints. It then uses a greedy first-fit algorithm that respects the constraints to provide placement locality (a tenant's VMs are placed close to VMs it communicates with) and places a tenant's VMs in the smallest subtree possible.
These approaches are similar to Quincy~\cite{quincy} and Firmament~\cite{firmament}, which also model the scheduling problem as a min-cost max-flow problem. Firmament's network-aware policy avoids bandwidth oversubscription at the end-host by incorporating applications network bandwidth demands into the flow graph. ~\cite{infocomvm} proposes an algorithm for traffic-aware VM placement that takes into account the traffic rates between VMs, and studies how different traffic patterns and \dc network architectures impact the algorithm's outcome. 

\subsubsection{Tail network latency guarantees} 

Tail latencies have been recognised as a source of significant performance degradation~\cite{Kapoor:2012, Dean:2013, detail, bobtail}. Several systems~\cite{silo, qjump, sncmeister} have been developed in response. Silo~\cite{silo} controls tenant's bandwidth to bound network queueing delay through packet pacing at the end-host. It then places VMs using a first-fit algorithm, while trying to place a tenant's VMs on the same server, in the same rack or further in the same pod, minimising the amount of network traffic that the core links have to cary. QJump~\cite{qjump} computes rate limits 
for classes of applications, ranging from latency-sensitive applications for which it offers strict latency guarantees to throughput-intensive ones for which latency can be variable. These systems provide worst-case latency guarantees. On the other hand, SNC-Meister~\cite{sncmeister} bases its design on the observation that tenants do not need worst-case guarantees, and that instead they require latency guarantees for lower percentiles, \eg, $99.9$\thss percentile. SNC-Meister leverages this observation to admit more tenants in a \dc while keeping their (lower percentile) latency guarantees.

\subsubsection{Application performance guarantees} All of the previous approaches looked at providing network bandwidth and (tail) latency guarantees for an application. NoMora~\cite{diana-nomora} seeks to provide application-level performance guarantees.
NoMora~\cite{diana-nomora} is a cluster scheduling architecture that places applications according to their predicted performance dependent upon network latency. Using application performance predictions dependent upon network latency and dynamic latency measurements obtained from PTPmesh~\cite{diana-mascots}, NoMora's cluster scheduling policy places applications to obtain the best performance under the current network conditions.
\subsection{Machine Learning (ML)-driven resource allocation and control}

Machine Learning for Systems promises to improve systems in an
automated manner, while getting rid of heuristics and manual tuning.
An early application of machine learning to systems is Internet traffic
classification~\cite{sigmetrics-moore}.

With the advent of machine learning, researchers have turned their attention 
to learning cluster scheduling policies through ML techniques rather than relying
on bespoke scheduling policies and heuristics, which are hard to devise and later tune
depending on the cluster workload. Leveraging cluster traces, the challenge in developing
a ML-driven cluster scheduler is adapting well-known ML algorithms to the cluster
scheduling problem. Out-of-the-box ML algorithms do not necessarily perform well,
and necessitate a new way to represent the arrival and scheduling of jobs in a cluster.

Several cluster schedulers used Deep Reinforcement Learning (RL) to determine a scheduling decision.
Decima~\cite{decima} is a cluster scheduler which trains a graph neural network using RL to make
decisions. Job Directed Acyclic Graphs (DAGs) are converted into vectors that serve as input to the graph neural network, whose
output is used in the policy network, which outputs scheduling decisions.
A scheduling agent observes the cluster state, takes a scheduling decision outputed by the network, and
receives a reward depending on an objective, such as minimizing job completion time or makespan. Decima outperforms
conventional cluster schedulers such as Tetris~\cite{tetris} and Graphene~\cite{graphene}.
Decima uses information regarding expected task duration, the number of tasks remaining at
each stage, and other cluster metrics. DeepRM~\cite{deeprm} and RLScheduler~\cite{rlscheduler} similarly use Deep Reinforcement Learning
for cluster scheduling. DeepRM~\cite{deeprm} schedules jobs using Reinforcement Learning with Deep Neural Networks (DNNs).
The state is represented by the current resource allocation, the resource profiles of the first
$M$ jobs waiting to be scheduled, and the number of jobs that are in the backlog (beyond the first
$M$ jobs). The objective is to minimize the average job completion time.
RLScheduler~\cite{rlscheduler} is a HPC batch job scheduler, and uses a kernel-based neural network structure 
that is insensitive to job order, which is the policy network. Based on the scheduling decisions
from the policy network, the rewards are used to train a value network to predict the reward
for a given job sequence. The input to the policy network is represented by job attributes
(job arrival time, number of requested processors, job estimated runtime, requested memory)
and available resources.	
Both Decima and RLScheduler introduce techniques to deal with variance in input in order
to ensure that the learning is not affected by different job arrival sequences.

Another area in which machine learning is being applied is scaling
automatically a workload in terms of resources~\cite{autopilot, cherrypick}.
Autopilot~\cite{autopilot} uses an exponentially-smoothing sliding window algorithm over historic
data for vertical scaling. An ML model is tuned for each job based on historical
data. CherryPick~\cite{cherrypick} builds performance models for applications
using Bayesian Optimization in order to find the optimal or near-optimal cloud resource allocation.

\textbf{\emph{Lessons Learned:}} There has been substantial research in the area of cluster 
scheduling, both conventional and network-aware. The emergence of ML-driven resource allocation and control
offers new opportunities for application performance guarantees in \dcs, since the
underlying systems that collect network monitoring data already exist. The next step is to
customize ML algorithms for the problems at hand, utilizing the rich network monitoring information.

\section{Future Directions and Conclusion}
\label{sec:future}

Providing network resource guarantees, and thus, predictable application
performance in \dcs remains challenging. Even though substantial progress
has been achieved in the last decade, several open challenges remain.

\subsection{Open-source network-related \dc traces}

Conducting \dc network-related research is challenging due to the lack of
up-to-date and comprehensive cluster workloads and network traffic traces.
The available cluster workloads do not include information related to required
network bandwidth and latency for cloud workloads.
In absence of this information, one way to mitigate this is to augment current available traces 
with typical cloud applications networking requirements~\cite{diana-nomora} or to generate
synthetic cloud workloads using ML techniques~\cite{sosp-huawei}. 
A similar challenge exists for network traffic. While companies have released network traffic traces,
these usually comprise sampled traffic. Traffic generator models for generating 
synthetic network traffic traces for \dcs that have similar complexity
have been proposed~\cite{sigmetrics-traces}. A collective effort is needed
to have an open-source collection of traces to aid in \dc network research,
similar to organisations that have organised to standardise benchmark suites 
in different areas.

\subsection{Network flow and cluster scheduling}

A substantial body of work deals with network flow scheduling, while
cluster scheduling is largely considered an orthogonal problem. The integration of
flow scheduling and cluster scheduling into a single monolithic
architecture would be interesting to explore. Usually, jobs
are placed in a \dc without taking into consideration network-related
resources. Then the network flow scheduling mechanism has to deal with the
network flows originating from the jobs. The research challenge is how do we achieve
an integration of the \dc management stack by finding an optimal
placement for a job both from the computational perspective and from the
network perspective. The multi-objective optimization that has to be solved in this case 
is complex.

\subsection{ML-driven network-aware cluster scheduling}

Applying machine learning algorithms to the cluster scheduling problem
is the obvious next step, since the \dc monitoring infrastructure is already in place, collecting
data that can be used to train new machine learning models to aid in resource allocation
and control. Recent work has already made progress in this area, but the current cluster scheduling ML algorithms
do not incorporate any network-related information about jobs.
A first step would be to use information such as network bandwidth requested by a job, 
job flow sizes, packet interarrival times, as additional inputs to these algorithms.
However, the current cluster scheduling ML algorithms are agnostic with regards to the network topology, which means that
scheduling decisions might be suboptimal and lead to network congestion. Designing ML algorithms that can encode
network topology and network state information will be important in automating network-aware \dc control. 
Furthermore, the objective of the ML-driven policy should take into account 
also flow completion time, not only objectives such as job completion time or makespan.

\vspace{0.3cm}
Our survey presents the reader with a unique perspective on measurement-based resource
allocation and control in \dcs. Network measurements offer the possibility of improving
cloud application performance both statically, by looking at previous application behaviour,
and dynamically, by ingesting new measurement data into the loop.

\ifCLASSOPTIONcaptionsoff
  \newpage
\fi



%

\bibliographystyle{IEEEtran}
\bibliography{references}

\begin{thebibliography}{100}
\providecommand{\url}[1]{#1}
\csname url@samestyle\endcsname
\providecommand{\newblock}{\relax}
\providecommand{\bibinfo}[2]{#2}
\providecommand{\BIBentrySTDinterwordspacing}{\spaceskip=0pt\relax}
\providecommand{\BIBentryALTinterwordstretchfactor}{4}
\providecommand{\BIBentryALTinterwordspacing}{\spaceskip=\fontdimen2\font plus
\BIBentryALTinterwordstretchfactor\fontdimen3\font minus
  \fontdimen4\font\relax}
\providecommand{\BIBforeignlanguage}[2]{{%
\expandafter\ifx\csname l@#1\endcsname\relax
\typeout{** WARNING: IEEEtran.bst: No hyphenation pattern has been}%
\typeout{** loaded for the language `#1'. Using the pattern for}%
\typeout{** the default language instead.}%
\else
\language=\csname l@#1\endcsname
\fi
#2}}
\providecommand{\BIBdecl}{\relax}
\BIBdecl

\bibitem{warehouse}
L.~A. Barroso and U.~Hoelzle, \emph{The Datacenter As a Computer: Designing
  Warehouse-Scale Machines, Third Edition}, 3rd~ed.\hskip 1em plus 0.5em minus
  0.4em\relax Morgan and Claypool Publishers, 2018.

\bibitem{diana-report}
\BIBentryALTinterwordspacing
D.~A. Popescu, N.~Zilberman, and A.~W. Moore, ``{Characterizing the impact of
  network latency on cloud-based applications' performance},'' University of
  Cambridge, Computer Laboratory, Tech. Rep. UCAM-CL-TR-914, Nov. 2017.
  [Online]. Available:
  \url{http://www.cl.cam.ac.uk/techreports/UCAM-CL-TR-914.pdf}
\BIBentrySTDinterwordspacing

\bibitem{diana-thesis}
\BIBentryALTinterwordspacing
D.~A. Popescu, ``{Latency-driven performance in data centres},'' University of
  Cambridge, Computer Laboratory, Tech. Rep. UCAM-CL-TR-937, Jun. 2019.
  [Online]. Available:
  \url{https://www.cl.cam.ac.uk/techreports/UCAM-CL-TR-937.pdf}
\BIBentrySTDinterwordspacing

\bibitem{mogul:2015}
\BIBentryALTinterwordspacing
J.~C. Mogul and R.~R. Kompella, ``Inferring the network latency requirements of
  cloud tenants,'' in \emph{Proceedings of the 15th USENIX Conference on Hot
  Topics in Operating Systems}, ser. HOTOS'15.\hskip 1em plus 0.5em minus
  0.4em\relax Berkeley, CA, USA: USENIX Association, 2015, pp. 24--24.
  [Online]. Available: \url{http://dl.acm.org/citation.cfm?id=2831090.2831114}
\BIBentrySTDinterwordspacing

\bibitem{Barroso}
\BIBentryALTinterwordspacing
L.~Barroso, M.~Marty, D.~Patterson, and P.~Ranganathan, ``Attack of the killer
  microseconds,'' \emph{Commun. ACM}, vol.~60, no.~4, pp. 48--54, Mar. 2017.
  [Online]. Available: \url{http://doi.acm.org/10.1145/3015146}
\BIBentrySTDinterwordspacing

\bibitem{pfabric}
\BIBentryALTinterwordspacing
M.~Alizadeh, S.~Yang, M.~Sharif, S.~Katti, N.~McKeown, B.~Prabhakar, and
  S.~Shenker, ``pfabric: Minimal near-optimal datacenter transport,'' in
  \emph{Proceedings of the ACM SIGCOMM 2013 Conference on SIGCOMM}, ser.
  SIGCOMM '13.\hskip 1em plus 0.5em minus 0.4em\relax New York, NY, USA: ACM,
  2013, pp. 435--446. [Online]. Available:
  \url{http://doi.acm.org/10.1145/2486001.2486031}
\BIBentrySTDinterwordspacing

\bibitem{fastpass}
\BIBentryALTinterwordspacing
J.~Perry, A.~Ousterhout, H.~Balakrishnan, D.~Shah, and H.~Fugal, ``Fastpass: A
  centralized "zero-queue" datacenter network,'' in \emph{Proceedings of the
  2014 ACM Conference on SIGCOMM}, ser. SIGCOMM '14.\hskip 1em plus 0.5em minus
  0.4em\relax New York, NY, USA: ACM, 2014, pp. 307--318. [Online]. Available:
  \url{http://doi.acm.org/10.1145/2619239.2626309}
\BIBentrySTDinterwordspacing

\bibitem{phost}
\BIBentryALTinterwordspacing
P.~X. Gao, A.~Narayan, G.~Kumar, R.~Agarwal, S.~Ratnasamy, and S.~Shenker,
  ``phost: Distributed near-optimal datacenter transport over commodity network
  fabric,'' in \emph{Proceedings of the 11th ACM Conference on Emerging
  Networking Experiments and Technologies}, ser. CoNEXT '15.\hskip 1em plus
  0.5em minus 0.4em\relax New York, NY, USA: ACM, 2015, pp. 1:1--1:12.
  [Online]. Available: \url{http://doi.acm.org/10.1145/2716281.2836086}
\BIBentrySTDinterwordspacing

\bibitem{qjump}
\BIBentryALTinterwordspacing
M.~P. Grosvenor, M.~Schwarzkopf, I.~Gog, R.~N.~M. Watson, A.~W. Moore, S.~Hand,
  and J.~Crowcroft, ``Queues don't matter when you can jump them!'' in
  \emph{Proceedings of the 12th USENIX Conference on Networked Systems Design
  and Implementation}, ser. NSDI'15.\hskip 1em plus 0.5em minus 0.4em\relax
  Berkeley, CA, USA: USENIX Association, 2015, pp. 1--14. [Online]. Available:
  \url{http://dl.acm.org/citation.cfm?id=2789770.2789771}
\BIBentrySTDinterwordspacing

\bibitem{ndp}
\BIBentryALTinterwordspacing
M.~Handley, C.~Raiciu, A.~Agache, A.~Voinescu, A.~W. Moore, G.~Antichi, and
  M.~W\'{o}jcik, ``Re-architecting datacenter networks and stacks for low
  latency and high performance,'' in \emph{Proceedings of the Conference of the
  ACM Special Interest Group on Data Communication}, ser. SIGCOMM '17.\hskip
  1em plus 0.5em minus 0.4em\relax New York, NY, USA: ACM, 2017, pp. 29--42.
  [Online]. Available: \url{http://doi.acm.org/10.1145/3098822.3098825}
\BIBentrySTDinterwordspacing

\bibitem{Hedera}
\BIBentryALTinterwordspacing
M.~Al-Fares, S.~Radhakrishnan, B.~Raghavan, N.~Huang, and A.~Vahdat, ``Hedera:
  Dynamic flow scheduling for data center networks,'' in \emph{Proceedings of
  the 7th USENIX Conference on Networked Systems Design and Implementation},
  ser. NSDI'10.\hskip 1em plus 0.5em minus 0.4em\relax Berkeley, CA, USA:
  USENIX Association, 2010, pp. 19--19. [Online]. Available:
  \url{http://dl.acm.org/citation.cfm?id=1855711.1855730}
\BIBentrySTDinterwordspacing

\bibitem{MicroTE}
\BIBentryALTinterwordspacing
T.~Benson, A.~Anand, A.~Akella, and M.~Zhang, ``{MicroTE: Fine Grained Traffic
  Engineering for Data Centers},'' in \emph{Proceedings of the Seventh
  COnference on Emerging Networking EXperiments and Technologies}, ser. CoNEXT
  '11.\hskip 1em plus 0.5em minus 0.4em\relax New York, NY, USA: ACM, 2011, pp.
  8:1--8:12. [Online]. Available:
  \url{http://doi.acm.org/10.1145/2079296.2079304}
\BIBentrySTDinterwordspacing

\bibitem{Conga}
\BIBentryALTinterwordspacing
M.~Alizadeh, T.~Edsall, S.~Dharmapurikar, R.~Vaidyanathan, K.~Chu,
  A.~Fingerhut, V.~T. Lam, F.~Matus, R.~Pan, N.~Yadav, and G.~Varghese,
  ``Conga: Distributed congestion-aware load balancing for datacenters,'' in
  \emph{Proceedings of the 2014 ACM Conference on SIGCOMM}, ser. SIGCOMM
  '14.\hskip 1em plus 0.5em minus 0.4em\relax New York, NY, USA: ACM, 2014, pp.
  503--514. [Online]. Available:
  \url{http://doi.acm.org/10.1145/2619239.2626316}
\BIBentrySTDinterwordspacing

\bibitem{vnet-pingmesh}
\BIBentryALTinterwordspacing
A.~Roy, D.~Bansal, D.~Brumley, H.~K. Chandrappa, P.~Sharma, R.~Tewari,
  B.~Arzani, and A.~C. Snoeren, ``Cloud datacenter sdn monitoring: Experiences
  and challenges,'' in \emph{Proceedings of the Internet Measurement Conference
  2018}, ser. IMC '18.\hskip 1em plus 0.5em minus 0.4em\relax New York, NY,
  USA: ACM, 2018, pp. 464--470. [Online]. Available:
  \url{http://doi.acm.org/10.1145/3278532.3278572}
\BIBentrySTDinterwordspacing

\bibitem{sdc-energy}
M.~Dayarathna, Y.~Wen, and R.~Fan, ``Data center energy consumption modeling: A
  survey,'' \emph{IEEE Communications Surveys Tutorials}, vol.~18, no.~1, pp.
  732--794, 2016.

\bibitem{sdc-virt}
M.~F. Bari, R.~Boutaba, R.~Esteves, L.~Z. Granville, M.~Podlesny, M.~G.
  Rabbani, Q.~Zhang, and M.~F. Zhani, ``Data center network virtualization: A
  survey,'' \emph{IEEE Communications Surveys Tutorials}, vol.~15, no.~2, pp.
  909--928, 2013.

\bibitem{sdc-lb}
J.~Zhang, F.~R. Yu, S.~Wang, T.~Huang, Z.~Liu, and Y.~Liu, ``Load balancing in
  data center networks: A survey,'' \emph{IEEE Communications Surveys
  Tutorials}, vol.~20, no.~3, pp. 2324--2352, 2018.

\bibitem{sdc-tc}
M.~Noormohammadpour and C.~S. Raghavendra, ``Datacenter traffic control:
  Understanding techniques and tradeoffs,'' \emph{IEEE Communications Surveys
  Tutorials}, vol.~20, no.~2, pp. 1492--1525, 2018.

\bibitem{sdc-io}
W.~Xia, P.~Zhao, Y.~Wen, and H.~Xie, ``A survey on data center networking
  (dcn): Infrastructure and operations,'' \emph{IEEE Communications Surveys
  Tutorials}, vol.~19, no.~1, pp. 640--656, 2017.

\bibitem{sdc-opt}
X.~Sun, N.~Ansari, and R.~Wang, ``Optimizing resource utilization of a data
  center,'' \emph{IEEE Communications Surveys Tutorials}, vol.~18, no.~4, pp.
  2822--2846, 2016.

\bibitem{sdc-cluster}
K.~Wang, Q.~Zhou, S.~Guo, and J.~Luo, ``Cluster frameworks for efficient
  scheduling and resource allocation in data center networks: A survey,''
  \emph{IEEE Communications Surveys Tutorials}, vol.~20, no.~4, pp. 3560--3580,
  2018.

\bibitem{Crovella}
M.~Crovella and B.~Krishnamurthy, \emph{Internet Measurement: Infrastructure,
  Traffic and Applications}.\hskip 1em plus 0.5em minus 0.4em\relax New York,
  NY, USA: John Wiley \& Sons, Inc., 2006.

\bibitem{snmp}
J.~D. Case, M.~Fedor, M.~L. Schoffstall, and J.~Davin, ``Simple network
  management protocol (snmp),'' United States, 1990.

\bibitem{netflow}
N.~Cisco, ``Netflow,''
  \url{https://www.cisco.com/c/en/us/products/ios-nx-os-software/ios-netflow/index.html},
  [Online; accessed December 2018].

\bibitem{sflow}
C.~sFlow, ``sflow,'' \url{https://sflow.org/}, [Online; accessed December
  2018].

\bibitem{psamp}
\BIBentryALTinterwordspacing
A.~Johnson, J.~Quittek, and B.~Claise, ``{Packet Sampling (PSAMP) Protocol
  Specifications},'' RFC 5476, Mar. 2009. [Online]. Available:
  \url{https://rfc-editor.org/rfc/rfc5476.txt}
\BIBentrySTDinterwordspacing

\bibitem{trajectory}
\BIBentryALTinterwordspacing
N.~G. Duffield and M.~Grossglauser, ``Trajectory sampling for direct traffic
  observation,'' \emph{IEEE/ACM Trans. Netw.}, vol.~9, no.~3, pp. 280--292,
  Jun. 2001. [Online]. Available: \url{http://dx.doi.org/10.1109/90.929851}
\BIBentrySTDinterwordspacing

\bibitem{samplednetflow}
S.~N. Cisco, ``Sampled netflow,''
  \url{https://www.cisco.com/c/en/us/td/docs/ios/12_0s/feature/guide/12s_sanf.html},
  [Online; accessed December 2018].

\bibitem{port-mirroring}
Cisco, ``{Catalyst Switched Port Analyzer (SPAN) Configuration Example},''
  \url{https://www.cisco.com/c/en/us/support/docs/switches/catalyst-6500-series-switches/10570-41.html},
  [Online; accessed December 2018].

\bibitem{iperf}
L.~B. N.~L. ESnet, ``{iPerf},'' \url{https://iperf.fr/}, [Online; accessed
  December 2018].

\bibitem{ipsla}
Cisco, ``Cisco ios ip slas configuration guide,''
  \url{http://www.cisco.com/c/en/us/td/docs/ios/12\_4/ip\_sla/
  configuration/guide/hsla\_c/hsoverv.html}, [Online; accessed December 2018].

\bibitem{ntp}
\BIBentryALTinterwordspacing
J.~Martin, J.~Burbank, W.~Kasch, and P.~D.~L. Mills, ``{Network Time Protocol
  Version 4: Protocol and Algorithms Specification},'' RFC 5905, Jun. 2010.
  [Online]. Available: \url{https://rfc-editor.org/rfc/rfc5905.txt}
\BIBentrySTDinterwordspacing

\bibitem{ptp}
IEEE, ``{IEEE 1588-2008 Precision Time Protocol},''
  \url{https://www.nist.gov/el/intelligent-systems-division-73500/introduction-ieee-1588},
  2008, [Online; accessed December 2018].

\bibitem{fattree}
\BIBentryALTinterwordspacing
M.~Al-Fares, A.~Loukissas, and A.~Vahdat, ``A scalable, commodity data center
  network architecture,'' in \emph{Proceedings of the ACM SIGCOMM 2008
  Conference on Data Communication}, ser. SIGCOMM '08.\hskip 1em plus 0.5em
  minus 0.4em\relax New York, NY, USA: ACM, 2008, pp. 63--74. [Online].
  Available: \url{http://doi.acm.org/10.1145/1402958.1402967}
\BIBentrySTDinterwordspacing

\bibitem{clos}
C.~Clos, ``A study of non-blocking switching networks,'' \emph{The Bell System
  Technical Journal}, vol.~32, no.~2, pp. 406--424, March 1953.

\bibitem{singh2015jupiter}
A.~Singh, J.~Ong, A.~Agarwal, G.~Anderson, A.~Armistead, R.~Bannon, S.~Boving,
  G.~Desai, B.~Felderman, P.~Germano \emph{et~al.}, ``Jupiter rising: A decade
  of clos topologies and centralized control in google's datacenter network,''
  \emph{ACM SIGCOMM Computer Communication Review}, vol.~45, no.~4, pp.
  183--197, 2015.

\bibitem{vl2}
\BIBentryALTinterwordspacing
A.~Greenberg, N.~Jain, S.~Kandula, C.~Kim, P.~Lahiri, D.~Maltz, P.~Patel, and
  S.~Sengupta, ``Vl2: A scalable and flexible data center network,'' in
  \emph{SIGCOMM}.\hskip 1em plus 0.5em minus 0.4em\relax Association for
  Computing Machinery, Inc., August 2009. [Online]. Available:
  \url{http://research.microsoft.com/apps/pubs/default.aspx?id=80693}
\BIBentrySTDinterwordspacing

\bibitem{facebook-dc}
A.~Andreyev, ``Introducing data center fabric, the next-generation facebook
  data center network,''
  \url{https://code.fb.com/production-engineering/introducing-data-center-fabric-the-next-generation-facebook-data-center-network/},
  2014, [Online; accessed December 2018].

\bibitem{ecmp}
C.~Hopps, ``Analysis of an equal-cost multi-path algorithm,'' RFC 2992, Tech.
  Rep., 2000.

\bibitem{Dcell}
\BIBentryALTinterwordspacing
C.~Guo, H.~Wu, K.~Tan, L.~Shi, Y.~Zhang, and S.~Lu, ``Dcell: A scalable and
  fault-tolerant network structure for data centers,'' in \emph{Proceedings of
  the ACM SIGCOMM 2008 Conference on Data Communication}, ser. SIGCOMM
  '08.\hskip 1em plus 0.5em minus 0.4em\relax New York, NY, USA: ACM, 2008, pp.
  75--86. [Online]. Available: \url{http://doi.acm.org/10.1145/1402958.1402968}
\BIBentrySTDinterwordspacing

\bibitem{BCube}
\BIBentryALTinterwordspacing
C.~Guo, G.~Lu, D.~Li, H.~Wu, X.~Zhang, Y.~Shi, C.~Tian, Y.~Zhang, and S.~Lu,
  ``Bcube: A high performance, server-centric network architecture for modular
  data centers,'' in \emph{Proceedings of the ACM SIGCOMM 2009 Conference on
  Data Communication}, ser. SIGCOMM '09.\hskip 1em plus 0.5em minus 0.4em\relax
  New York, NY, USA: ACM, 2009, pp. 63--74. [Online]. Available:
  \url{http://doi.acm.org/10.1145/1592568.1592577}
\BIBentrySTDinterwordspacing

\bibitem{CamCube}
\BIBentryALTinterwordspacing
H.~Abu-Libdeh, P.~Costa, A.~Rowstron, G.~O'Shea, and A.~Donnelly, ``Symbiotic
  routing in future data centers,'' \emph{SIGCOMM Comput. Commun. Rev.},
  vol.~40, no.~4, pp. 51--62, Aug. 2010. [Online]. Available:
  \url{http://doi.acm.org/10.1145/1851275.1851191}
\BIBentrySTDinterwordspacing

\bibitem{Jellyfish}
\BIBentryALTinterwordspacing
A.~Singla, C.-Y. Hong, L.~Popa, and P.~B. Godfrey, ``Jellyfish: Networking data
  centers randomly,'' in \emph{Proceedings of the 9th USENIX Conference on
  Networked Systems Design and Implementation}, ser. NSDI'12.\hskip 1em plus
  0.5em minus 0.4em\relax Berkeley, CA, USA: USENIX Association, 2012, pp.
  17--17. [Online]. Available:
  \url{http://dl.acm.org/citation.cfm?id=2228298.2228322}
\BIBentrySTDinterwordspacing

\bibitem{xpander}
\BIBentryALTinterwordspacing
A.~Valadarsky, G.~Shahaf, M.~Dinitz, and M.~Schapira, ``Xpander: Towards
  optimal-performance datacenters,'' in \emph{Proceedings of the 12th
  International on Conference on Emerging Networking EXperiments and
  Technologies}, ser. CoNEXT '16.\hskip 1em plus 0.5em minus 0.4em\relax New
  York, NY, USA: ACM, 2016, pp. 205--219. [Online]. Available:
  \url{http://doi.acm.org/10.1145/2999572.2999580}
\BIBentrySTDinterwordspacing

\bibitem{gao2016}
\BIBentryALTinterwordspacing
P.~X. Gao, A.~Narayan, S.~Karandikar, J.~Carreira, S.~Han, R.~Agarwal,
  S.~Ratnasamy, and S.~Shenker, ``Network requirements for resource
  disaggregation,'' in \emph{Proceedings of the 12th USENIX Conference on
  Operating Systems Design and Implementation}, ser. OSDI'16.\hskip 1em plus
  0.5em minus 0.4em\relax Berkeley, CA, USA: USENIX Association, 2016, pp.
  249--264. [Online]. Available:
  \url{http://dl.acm.org/citation.cfm?id=3026877.3026897}
\BIBentrySTDinterwordspacing

\bibitem{legoos}
\BIBentryALTinterwordspacing
Y.~Shan, Y.~Huang, Y.~Chen, and Y.~Zhang, ``Legoos: A disseminated, distributed
  {OS} for hardware resource disaggregation,'' in \emph{13th {USENIX} Symposium
  on Operating Systems Design and Implementation ({OSDI} 18)}.\hskip 1em plus
  0.5em minus 0.4em\relax Carlsbad, CA: {USENIX} Association, 2018, pp. 69--87.
  [Online]. Available:
  \url{https://www.usenix.org/conference/osdi18/presentation/shan}
\BIBentrySTDinterwordspacing

\bibitem{intel-disaggregated}
Intel, ``{Intel Disaggregated Servers Drive Data Center Efficiency and
  Innovation},''
  \url{https://www.intel.co.uk/content/www/uk/en/it-management/intel-it-best-practices/disaggregated-server-architecture-drives-data-center-efficiency-paper.html},
  2018, [Online; accessed December 2018].

\bibitem{arp}
D.~C. Plummer, ``Ethernet address resolution protocol: Or converting network
  protocol addresses to 48.bit ethernet address for transmission on ethernet
  hardware,'' United States, 1982.

\bibitem{portland}
\BIBentryALTinterwordspacing
R.~Niranjan~Mysore, A.~Pamboris, N.~Farrington, N.~Huang, P.~Miri,
  S.~Radhakrishnan, V.~Subramanya, and A.~Vahdat, ``Portland: A scalable
  fault-tolerant layer 2 data center network fabric,'' in \emph{Proceedings of
  the ACM SIGCOMM 2009 Conference on Data Communication}, ser. SIGCOMM
  '09.\hskip 1em plus 0.5em minus 0.4em\relax New York, NY, USA: ACM, 2009, pp.
  39--50. [Online]. Available: \url{http://doi.acm.org/10.1145/1592568.1592575}
\BIBentrySTDinterwordspacing

\bibitem{bgp}
Y.~Rekhter and T.~Li, ``A border gateway protocol 4 (bgp-4),'' United States,
  1995.

\bibitem{alizadeh2010dctcp}
\BIBentryALTinterwordspacing
M.~Alizadeh, A.~Greenberg, D.~A. Maltz, J.~Padhye, P.~Patel, B.~Prabhakar,
  S.~Sengupta, and M.~Sridharan, ``Data center tcp (dctcp),'' in
  \emph{Proceedings of the ACM SIGCOMM 2010 Conference}, ser. SIGCOMM
  '10.\hskip 1em plus 0.5em minus 0.4em\relax New York, NY, USA: ACM, 2010, pp.
  63--74. [Online]. Available: \url{http://doi.acm.org/10.1145/1851182.1851192}
\BIBentrySTDinterwordspacing

\bibitem{hull}
\BIBentryALTinterwordspacing
M.~Alizadeh, A.~Kabbani, T.~Edsall, B.~Prabhakar, A.~Vahdat, and M.~Yasuda,
  ``Less is more: Trading a little bandwidth for ultra-low latency in the data
  center,'' in \emph{Proceedings of the 9th USENIX Conference on Networked
  Systems Design and Implementation}, ser. NSDI'12.\hskip 1em plus 0.5em minus
  0.4em\relax Berkeley, CA, USA: USENIX Association, 2012, pp. 19--19.
  [Online]. Available: \url{http://dl.acm.org/citation.cfm?id=2228298.2228324}
\BIBentrySTDinterwordspacing

\bibitem{d2tcp}
\BIBentryALTinterwordspacing
B.~Vamanan, J.~Hasan, and T.~Vijaykumar, ``Deadline-aware datacenter tcp
  (d2tcp),'' in \emph{Proceedings of the ACM SIGCOMM 2012 Conference on
  Applications, Technologies, Architectures, and Protocols for Computer
  Communication}, ser. SIGCOMM '12.\hskip 1em plus 0.5em minus 0.4em\relax New
  York, NY, USA: ACM, 2012, pp. 115--126. [Online]. Available:
  \url{http://doi.acm.org/10.1145/2342356.2342388}
\BIBentrySTDinterwordspacing

\bibitem{ptpd}
PTPd, ``{PTP daemon},'' \url{https://github.com/ptpd/ptpd}, 2018, [Online;
  accessed December 2018].

\bibitem{chubby}
M.~Burrows, ``The chubby lock service for loosely-coupled distributed
  systems,'' in \emph{7th {USENIX} Symposium on Operating Systems Design and
  Implementation (OSDI)}, 2006.

\bibitem{borg}
\BIBentryALTinterwordspacing
A.~Verma, L.~Pedrosa, M.~Korupolu, D.~Oppenheimer, E.~Tune, and J.~Wilkes,
  ``Large-scale cluster management at google with borg,'' in \emph{Proceedings
  of the Tenth European Conference on Computer Systems}, ser. EuroSys
  '15.\hskip 1em plus 0.5em minus 0.4em\relax New York, NY, USA: ACM, 2015, pp.
  18:1--18:17. [Online]. Available:
  \url{http://doi.acm.org/10.1145/2741948.2741964}
\BIBentrySTDinterwordspacing

\bibitem{GFS}
S.~Ghemawat, H.~Gobioff, and S.-T. Leung, ``The google file system,'' in
  \emph{Proceedings of the 19th ACM Symposium on Operating Systems Principles},
  Bolton Landing, NY, 2003, pp. 20--43.

\bibitem{spanner}
\BIBentryALTinterwordspacing
J.~C. Corbett, J.~Dean, M.~Epstein, A.~Fikes, C.~Frost, J.~J. Furman,
  S.~Ghemawat, A.~Gubarev, C.~Heiser, P.~Hochschild, W.~Hsieh, S.~Kanthak,
  E.~Kogan, H.~Li, A.~Lloyd, S.~Melnik, D.~Mwaura, D.~Nagle, S.~Quinlan,
  R.~Rao, L.~Rolig, Y.~Saito, M.~Szymaniak, C.~Taylor, R.~Wang, and
  D.~Woodford, ``Spanner: Google's globally-distributed database,'' in
  \emph{Proceedings of the 10th USENIX Conference on Operating Systems Design
  and Implementation}, ser. OSDI'12.\hskip 1em plus 0.5em minus 0.4em\relax
  Berkeley, CA, USA: USENIX Association, 2012, pp. 251--264. [Online].
  Available: \url{http://dl.acm.org/citation.cfm?id=2387880.2387905}
\BIBentrySTDinterwordspacing

\bibitem{memcached}
Memcached, ``{Memcached},'' \url{https://memcached.org/}, 2018, [Online;
  accessed December 2018].

\bibitem{MapReduce}
J.~Dean and S.~Ghemawat, ``Mapreduce: Simplified data processing on large
  clusters,'' in \emph{OSDI'04: Sixth Symposium on Operating System Design and
  Implementation}, San Francisco, CA, 2004, pp. 137--150.

\bibitem{giraph}
A.~Giraph, ``{Apache Giraph},'' \url{http://giraph.apache.org/}, [Online;
  accessed December 2018].

\bibitem{pregel}
\BIBentryALTinterwordspacing
G.~Malewicz, M.~H. Austern, A.~J. Bik, J.~C. Dehnert, I.~Horn, N.~Leiser, and
  G.~Czajkowski, ``Pregel: A system for large-scale graph processing,'' in
  \emph{Proceedings of the 2010 ACM SIGMOD International Conference on
  Management of Data}, ser. SIGMOD '10.\hskip 1em plus 0.5em minus 0.4em\relax
  New York, NY, USA: ACM, 2010, pp. 135--146. [Online]. Available:
  \url{http://doi.acm.org/10.1145/1807167.1807184}
\BIBentrySTDinterwordspacing

\bibitem{storm}
A.~Storm, ``{Apache Storm},'' \url{http://storm.apache.org/}, [Online; accessed
  December 2018].

\bibitem{tensorflow}
\BIBentryALTinterwordspacing
M.~Abadi, P.~Barham, J.~Chen, Z.~Chen, A.~Davis, J.~Dean, M.~Devin,
  S.~Ghemawat, G.~Irving, M.~Isard, M.~Kudlur, J.~Levenberg, R.~Monga,
  S.~Moore, D.~G. Murray, B.~Steiner, P.~Tucker, V.~Vasudevan, P.~Warden,
  M.~Wicke, Y.~Yu, and X.~Zheng, ``Tensorflow: A system for large-scale machine
  learning,'' in \emph{Proceedings of the 12th USENIX Conference on Operating
  Systems Design and Implementation}, ser. OSDI'16.\hskip 1em plus 0.5em minus
  0.4em\relax Berkeley, CA, USA: USENIX Association, 2016, pp. 265--283.
  [Online]. Available: \url{http://dl.acm.org/citation.cfm?id=3026877.3026899}
\BIBentrySTDinterwordspacing

\bibitem{apache-web-server}
A.~H.~S. Project, ``{Apache HTTP Server Project},''
  \url{https://www.mysql.com/}, 2018, [Online; accessed December 2018].

\bibitem{coflow}
\BIBentryALTinterwordspacing
M.~Chowdhury and I.~Stoica, ``Coflow: A networking abstraction for cluster
  applications,'' in \emph{Proceedings of the 11th ACM Workshop on Hot Topics
  in Networks}, ser. HotNets-XI.\hskip 1em plus 0.5em minus 0.4em\relax New
  York, NY, USA: ACM, 2012, pp. 31--36. [Online]. Available:
  \url{http://doi.acm.org/10.1145/2390231.2390237}
\BIBentrySTDinterwordspacing

\bibitem{Kapoor:2012}
\BIBentryALTinterwordspacing
R.~Kapoor, G.~Porter, M.~Tewari, G.~M. Voelker, and A.~Vahdat, ``Chronos:
  Predictable low latency for data center applications,'' in \emph{Proceedings
  of the Third ACM Symposium on Cloud Computing}, ser. SoCC '12.\hskip 1em plus
  0.5em minus 0.4em\relax New York, NY, USA: ACM, 2012, pp. 9:1--9:14.
  [Online]. Available: \url{http://doi.acm.org/10.1145/2391229.2391238}
\BIBentrySTDinterwordspacing

\bibitem{parameter-server}
\BIBentryALTinterwordspacing
M.~Li, D.~G. Andersen, J.~W. Park, A.~J. Smola, A.~Ahmed, V.~Josifovski,
  J.~Long, E.~J. Shekita, and B.-Y. Su, ``Scaling distributed machine learning
  with the parameter server,'' in \emph{11th {USENIX} Symposium on Operating
  Systems Design and Implementation ({OSDI} 14)}.\hskip 1em plus 0.5em minus
  0.4em\relax Broomfield, CO: {USENIX} Association, 2014, pp. 583--598.
  [Online]. Available:
  \url{https://www.usenix.org/conference/osdi14/technical-sessions/presentation/li_mu}
\BIBentrySTDinterwordspacing

\bibitem{spark-rdd}
\BIBentryALTinterwordspacing
M.~Zaharia, M.~Chowdhury, T.~Das, A.~Dave, J.~Ma, M.~McCauly, M.~J. Franklin,
  S.~Shenker, and I.~Stoica, ``Resilient distributed datasets: A fault-tolerant
  abstraction for in-memory cluster computing,'' in \emph{Presented as part of
  the 9th {USENIX} Symposium on Networked Systems Design and Implementation
  ({NSDI} 12)}.\hskip 1em plus 0.5em minus 0.4em\relax San Jose, CA: {USENIX},
  2012, pp. 15--28. [Online]. Available:
  \url{https://www.usenix.org/conference/nsdi12/technical-sessions/presentation/zaharia}
\BIBentrySTDinterwordspacing

\bibitem{ml-at-facebook}
K.~Hazelwood, S.~Bird, D.~Brooks, S.~Chintala, U.~Diril, D.~Dzhulgakov,
  M.~Fawzy, B.~Jia, Y.~Jia, A.~Kalro, J.~Law, K.~Lee, J.~Lu, P.~Noordhuis,
  M.~Smelyanskiy, L.~Xiong, and X.~Wang, ``Applied machine learning at
  facebook: A datacenter infrastructure perspective,'' in \emph{2018 IEEE
  International Symposium on High Performance Computer Architecture (HPCA)},
  Feb 2018, pp. 620--629.

\bibitem{Kim:2016}
\BIBentryALTinterwordspacing
J.~K. Kim, Q.~Ho, S.~Lee, X.~Zheng, W.~Dai, G.~A. Gibson, and E.~P. Xing,
  ``Strads: A distributed framework for scheduled model parallel machine
  learning,'' in \emph{Proceedings of the Eleventh European Conference on
  Computer Systems}, ser. EuroSys '16.\hskip 1em plus 0.5em minus 0.4em\relax
  New York, NY, USA: ACM, 2016, pp. 5:1--5:16. [Online]. Available:
  \url{http://doi.acm.org/10.1145/2901318.2901331}
\BIBentrySTDinterwordspacing

\bibitem{petuum}
E.~P. Xing, Q.~Ho, W.~Dai, J.~K. Kim, J.~Wei, S.~Lee, X.~Zheng, P.~Xie,
  A.~Kumar, and Y.~Yu, ``Petuum: A new platform for distributed machine
  learning on big data,'' \emph{IEEE Transactions on Big Data}, vol.~1, no.~2,
  pp. 49--67, June 2015.

\bibitem{Atikoglu:2012}
\BIBentryALTinterwordspacing
B.~Atikoglu, Y.~Xu, E.~Frachtenberg, S.~Jiang, and M.~Paleczny, ``Workload
  analysis of a large-scale key-value store,'' in \emph{Proceedings of the 12th
  ACM SIGMETRICS/PERFORMANCE Joint International Conference on Measurement and
  Modeling of Computer Systems}, ser. SIGMETRICS '12.\hskip 1em plus 0.5em
  minus 0.4em\relax New York, NY, USA: ACM, 2012, pp. 53--64. [Online].
  Available: \url{http://doi.acm.org/10.1145/2254756.2254766}
\BIBentrySTDinterwordspacing

\bibitem{Leverich:2014}
\BIBentryALTinterwordspacing
J.~Leverich and C.~Kozyrakis, ``Reconciling high server utilization and
  sub-millisecond quality-of-service,'' in \emph{Proceedings of the Ninth
  European Conference on Computer Systems}, ser. EuroSys '14.\hskip 1em plus
  0.5em minus 0.4em\relax New York, NY, USA: ACM, 2014, pp. 4:1--4:14.
  [Online]. Available: \url{http://doi.acm.org/10.1145/2592798.2592821}
\BIBentrySTDinterwordspacing

\bibitem{mapreduceworkload}
\BIBentryALTinterwordspacing
Y.~Chen, S.~Alspaugh, and R.~Katz, ``Interactive analytical processing in big
  data systems: A cross-industry study of mapreduce workloads,'' \emph{Proc.
  VLDB Endow.}, vol.~5, no.~12, pp. 1802--1813, Aug. 2012. [Online]. Available:
  \url{http://dx.doi.org/10.14778/2367502.2367519}
\BIBentrySTDinterwordspacing

\bibitem{mnist}
Y.~LeCun and C.~Cortes, ``{MNIST} handwritten digit database,''
  \url{http://yann.lecun.com/exdb/mnist/}, 2010, [Online; accessed December
  2018].

\bibitem{imagenet}
J.~Deng, W.~Dong, R.~Socher, L.-J. Li, K.~Li, and L.~Fei-Fei, ``{ImageNet: A
  Large-Scale Hierarchical Image Database},'' in \emph{CVPR09}, 2009.

\bibitem{dctraffic}
\BIBentryALTinterwordspacing
S.~Kandula, S.~Sengupta, A.~Greenberg, P.~Patel, and R.~Chaiken, ``The nature
  of data center traffic: Measurements \& analysis,'' in \emph{Proceedings of
  the 9th ACM SIGCOMM Conference on Internet Measurement Conference}, ser. IMC
  '09.\hskip 1em plus 0.5em minus 0.4em\relax New York, NY, USA: ACM, 2009, pp.
  202--208. [Online]. Available:
  \url{http://doi.acm.org/10.1145/1644893.1644918}
\BIBentrySTDinterwordspacing

\bibitem{benson}
\BIBentryALTinterwordspacing
T.~Benson, A.~Akella, and D.~A. Maltz, ``Network traffic characteristics of
  data centers in the wild,'' in \emph{Proceedings of the 10th ACM SIGCOMM
  Conference on Internet Measurement}, ser. IMC '10.\hskip 1em plus 0.5em minus
  0.4em\relax New York, NY, USA: ACM, 2010, pp. 267--280. [Online]. Available:
  \url{http://doi.acm.org/10.1145/1879141.1879175}
\BIBentrySTDinterwordspacing

\bibitem{roy2015}
\BIBentryALTinterwordspacing
A.~Roy, H.~Zeng, J.~Bagga, G.~Porter, and A.~C. Snoeren, ``Inside the social
  network's (datacenter) network,'' in \emph{Proceedings of the 2015 ACM
  Conference on Special Interest Group on Data Communication}, ser. SIGCOMM
  '15.\hskip 1em plus 0.5em minus 0.4em\relax New York, NY, USA: ACM, 2015, pp.
  123--137. [Online]. Available:
  \url{http://doi.acm.org/10.1145/2785956.2787472}
\BIBentrySTDinterwordspacing

\bibitem{infocomvm}
\BIBentryALTinterwordspacing
X.~Meng, V.~Pappas, and L.~Zhang, ``{Improving the Scalability of Data Center
  Networks with Traffic-aware Virtual Machine Placement},'' in
  \emph{Proceedings of the 29th Conference on Information Communications}, ser.
  INFOCOM'10.\hskip 1em plus 0.5em minus 0.4em\relax Piscataway, NJ, USA: IEEE
  Press, 2010, pp. 1154--1162. [Online]. Available:
  \url{http://dl.acm.org/citation.cfm?id=1833515.1833690}
\BIBentrySTDinterwordspacing

\bibitem{orchestra}
\BIBentryALTinterwordspacing
M.~Chowdhury, M.~Zaharia, J.~Ma, M.~I. Jordan, and I.~Stoica, ``{Managing Data
  Transfers in Computer Clusters with Orchestra},'' in \emph{Proceedings of the
  ACM SIGCOMM 2011 Conference}, ser. SIGCOMM '11.\hskip 1em plus 0.5em minus
  0.4em\relax New York, NY, USA: ACM, 2011, pp. 98--109. [Online]. Available:
  \url{http://doi.acm.org/10.1145/2018436.2018448}
\BIBentrySTDinterwordspacing

\bibitem{augmenting}
\BIBentryALTinterwordspacing
D.~Halperin, S.~Kandula, J.~Padhye, P.~Bahl, and D.~Wetherall, ``{Augmenting
  Data Center Networks with Multi-gigabit Wireless Links},'' in
  \emph{Proceedings of the ACM SIGCOMM 2011 Conference}, ser. SIGCOMM
  '11.\hskip 1em plus 0.5em minus 0.4em\relax New York, NY, USA: ACM, 2011, pp.
  38--49. [Online]. Available: \url{http://doi.acm.org/10.1145/2018436.2018442}
\BIBentrySTDinterwordspacing

\bibitem{google-workload}
\BIBentryALTinterwordspacing
C.~Reiss, A.~Tumanov, G.~R. Ganger, R.~H. Katz, and M.~A. Kozuch,
  ``Heterogeneity and dynamicity of clouds at scale: Google trace analysis,''
  in \emph{Proceedings of the Third ACM Symposium on Cloud Computing}, ser.
  SoCC '12.\hskip 1em plus 0.5em minus 0.4em\relax New York, NY, USA: ACM,
  2012, pp. 7:1--7:13. [Online]. Available:
  \url{http://doi.acm.org/10.1145/2391229.2391236}
\BIBentrySTDinterwordspacing

\bibitem{cortez2017}
\BIBentryALTinterwordspacing
E.~Cortez, A.~Bonde, A.~Muzio, M.~Russinovich, M.~Fontoura, and R.~Bianchini,
  ``Resource central: Understanding and predicting workloads for improved
  resource management in large cloud platforms,'' in \emph{Proceedings of the
  26th Symposium on Operating Systems Principles}, ser. SOSP '17.\hskip 1em
  plus 0.5em minus 0.4em\relax New York, NY, USA: ACM, 2017, pp. 153--167.
  [Online]. Available: \url{http://doi.acm.org/10.1145/3132747.3132772}
\BIBentrySTDinterwordspacing

\bibitem{cluster-workloads}
\BIBentryALTinterwordspacing
G.~Amvrosiadis, J.~W. Park, G.~R. Ganger, G.~A. Gibson, E.~Baseman, and
  N.~DeBardeleben, ``On the diversity of cluster workloads and its impact on
  research results,'' in \emph{2018 {USENIX} Annual Technical Conference
  ({USENIX} {ATC} 18)}.\hskip 1em plus 0.5em minus 0.4em\relax Boston, MA:
  {USENIX} Association, 2018, pp. 533--546. [Online]. Available:
  \url{https://www.usenix.org/conference/atc18/presentation/amvrosiadis}
\BIBentrySTDinterwordspacing

\bibitem{borg-next}
\BIBentryALTinterwordspacing
M.~Tirmazi, A.~Barker, N.~Deng, M.~E. Haque, Z.~G. Qin, S.~Hand,
  M.~Harchol-Balter, and J.~Wilkes, ``Borg: The next generation,'' in
  \emph{Proceedings of the Fifteenth European Conference on Computer Systems},
  ser. EuroSys '20.\hskip 1em plus 0.5em minus 0.4em\relax New York, NY, USA:
  Association for Computing Machinery, 2020. [Online]. Available:
  \url{https://doi.org/10.1145/3342195.3387517}
\BIBentrySTDinterwordspacing

\bibitem{OpenFlow}
\BIBentryALTinterwordspacing
N.~McKeown, T.~Anderson, H.~Balakrishnan, G.~Parulkar, L.~Peterson, J.~Rexford,
  S.~Shenker, and J.~Turner, ``Openflow: Enabling innovation in campus
  networks,'' \emph{SIGCOMM Comput. Commun. Rev.}, vol.~38, no.~2, pp. 69--74,
  Mar. 2008. [Online]. Available:
  \url{http://doi.acm.org/10.1145/1355734.1355746}
\BIBentrySTDinterwordspacing

\bibitem{OpenTM}
\BIBentryALTinterwordspacing
A.~Tootoonchian, M.~Ghobadi, and Y.~Ganjali, ``Opentm: Traffic matrix estimator
  for openflow networks,'' in \emph{Proceedings of the 11th International
  Conference on Passive and Active Measurement}, ser. PAM'10.\hskip 1em plus
  0.5em minus 0.4em\relax Berlin, Heidelberg: Springer-Verlag, 2010, pp.
  201--210. [Online]. Available:
  \url{http://dl.acm.org/citation.cfm?id=1889324.1889345}
\BIBentrySTDinterwordspacing

\bibitem{FlowSense}
\BIBentryALTinterwordspacing
C.~Yu, C.~Lumezanu, Y.~Zhang, V.~Singh, G.~Jiang, and H.~V. Madhyastha,
  ``Flowsense: Monitoring network utilization with zero measurement cost,'' in
  \emph{Proceedings of the 14th International Conference on Passive and Active
  Measurement}, ser. PAM'13.\hskip 1em plus 0.5em minus 0.4em\relax Berlin,
  Heidelberg: Springer-Verlag, 2013, pp. 31--41. [Online]. Available:
  \url{http://dx.doi.org/10.1007/978-3-642-36516-4_4}
\BIBentrySTDinterwordspacing

\bibitem{PayLess}
S.~Chowdhury, M.~Bari, R.~Ahmed, and R.~Boutaba, ``{PayLess}: A low cost
  network monitoring framework for software defined networks,'' in \emph{2014
  {IEEE} Network Operations and Management Symposium ({NOMS})}, May 2014, pp.
  1--9.

\bibitem{dream}
\BIBentryALTinterwordspacing
M.~Moshref, M.~Yu, R.~Govindan, and A.~Vahdat, ``Dream: Dynamic resource
  allocation for software-defined measurement,'' in \emph{Proceedings of the
  2014 ACM Conference on SIGCOMM}, ser. SIGCOMM '14.\hskip 1em plus 0.5em minus
  0.4em\relax New York, NY, USA: ACM, 2014, pp. 419--430. [Online]. Available:
  \url{http://doi.acm.org/10.1145/2619239.2626291}
\BIBentrySTDinterwordspacing

\bibitem{opensketch}
\BIBentryALTinterwordspacing
M.~Yu, L.~Jose, and R.~Miao, ``Software defined traffic measurement with
  opensketch,'' in \emph{Proceedings of the 10th USENIX Conference on Networked
  Systems Design and Implementation}, ser. nsdi'13.\hskip 1em plus 0.5em minus
  0.4em\relax Berkeley, CA, USA: USENIX Association, 2013, pp. 29--42.
  [Online]. Available: \url{http://dl.acm.org/citation.cfm?id=2482626.2482631}
\BIBentrySTDinterwordspacing

\bibitem{slam}
\BIBentryALTinterwordspacing
C.~Yu, C.~Lumezanu, A.~Sharma, Q.~Xu, G.~Jiang, and H.~V. Madhyastha,
  \emph{Software-Defined Latency Monitoring in Data Center Networks}.\hskip 1em
  plus 0.5em minus 0.4em\relax Cham: Springer International Publishing, 2015,
  pp. 360--372. [Online]. Available:
  \url{http://dx.doi.org/10.1007/978-3-319-15509-8\_27}
\BIBentrySTDinterwordspacing

\bibitem{OpenSample}
J.~Suh, T.~T. Kwon, C.~Dixon, W.~Felter, and J.~Carter, ``{OpenSample}: A
  low-latency, sampling-based measurement platform for commodity {SDN},'' in
  \emph{2014 {IEEE} 34th International Conference on Distributed Computing
  Systems ({ICDCS})}, Jun. 2014, pp. 228--237.

\bibitem{Planck}
\BIBentryALTinterwordspacing
J.~Rasley, B.~Stephens, C.~Dixon, E.~Rozner, W.~Felter, K.~Agarwal, J.~Carter,
  and R.~Fonseca, ``Planck: Millisecond-scale monitoring and control for
  commodity networks,'' in \emph{Proceedings of the 2014 ACM Conference on
  SIGCOMM}, ser. SIGCOMM '14.\hskip 1em plus 0.5em minus 0.4em\relax New York,
  NY, USA: ACM, 2014, pp. 407--418. [Online]. Available:
  \url{http://doi.acm.org/10.1145/2619239.2626310}
\BIBentrySTDinterwordspacing

\bibitem{netnorad}
A.~Adams, P.~Lapukhov, and J.~H. Zeng, ``{NetNORAD: Troubleshooting networks
  via end-to-end probing},''
  \url{https://code.facebook.com/posts/1534350660228025/
  netnorad-troubleshooting-networks-via-end-to-end-probing/}, 2016, [Online;
  accessed December 2018].

\bibitem{everflow}
\BIBentryALTinterwordspacing
Y.~Zhu, N.~Kang, J.~Cao, A.~Greenberg, G.~Lu, R.~Mahajan, D.~Maltz, L.~Yuan,
  M.~Zhang, B.~Y. Zhao, and H.~Zheng, ``Packet-level telemetry in large
  datacenter networks,'' in \emph{Proceedings of the 2015 ACM Conference on
  Special Interest Group on Data Communication}, ser. SIGCOMM '15.\hskip 1em
  plus 0.5em minus 0.4em\relax New York, NY, USA: ACM, 2015, pp. 479--491.
  [Online]. Available: \url{http://doi.acm.org/10.1145/2785956.2787483}
\BIBentrySTDinterwordspacing

\bibitem{pingmesh}
\BIBentryALTinterwordspacing
C.~Guo, L.~Yuan, D.~Xiang, Y.~Dang, R.~Huang, D.~Maltz, Z.~Liu, V.~Wang,
  B.~Pang, H.~Chen, Z.-W. Lin, and V.~Kurien, ``Pingmesh: A large-scale system
  for data center network latency measurement and analysis,'' in
  \emph{Proceedings of the 2015 ACM Conference on Special Interest Group on
  Data Communication}, ser. SIGCOMM '15.\hskip 1em plus 0.5em minus 0.4em\relax
  New York, NY, USA: ACM, 2015, pp. 139--152. [Online]. Available:
  \url{http://doi.acm.org/10.1145/2785956.2787496}
\BIBentrySTDinterwordspacing

\bibitem{diana-mascots}
D.~A. Popescu and A.~W. Moore, ``Ptpmesh: Data center network latency
  measurements using ptp,'' in \emph{2017 IEEE 25th International Symposium on
  Modeling, Analysis, and Simulation of Computer and Telecommunication Systems
  (MASCOTS)}, Sept 2017, pp. 73--79.

\bibitem{diana-tma}
{D. A. Popescu and A. W. Moore}, ``{A First Look at Data Center Network
  Condition Through The Eyes of PTPmesh},'' in \emph{2018 Network Traffic
  Measurement and Analysis Conference (TMA)}, June 2018, pp. 1--8.

\bibitem{diana-tnsm}
D.~A. Popescu and A.~W. Moore, ``{Measuring Network Conditions in Data Centers
  Using the Precision Time Protocol},'' \emph{IEEE Transactions on Network and
  Service Management}, vol.~18, no.~3, pp. 3753--3770, 2021.

\bibitem{univmon}
\BIBentryALTinterwordspacing
Z.~Liu, A.~Manousis, G.~Vorsanger, V.~Sekar, and V.~Braverman, ``One sketch to
  rule them all: Rethinking network flow monitoring with univmon,'' in
  \emph{Proceedings of the 2016 ACM SIGCOMM Conference}, ser. SIGCOMM
  '16.\hskip 1em plus 0.5em minus 0.4em\relax New York, NY, USA: ACM, 2016, pp.
  101--114. [Online]. Available:
  \url{http://doi.acm.org/10.1145/2934872.2934906}
\BIBentrySTDinterwordspacing

\bibitem{elasticsketch}
\BIBentryALTinterwordspacing
T.~Yang, J.~Jiang, P.~Liu, Q.~Huang, J.~Gong, Y.~Zhou, R.~Miao, X.~Li, and
  S.~Uhlig, ``Elastic sketch: Adaptive and fast network-wide measurements,'' in
  \emph{Proceedings of the 2018 Conference of the ACM Special Interest Group on
  Data Communication}, ser. SIGCOMM '18.\hskip 1em plus 0.5em minus 0.4em\relax
  New York, NY, USA: ACM, 2018, pp. 561--575. [Online]. Available:
  \url{http://doi.acm.org/10.1145/3230543.3230544}
\BIBentrySTDinterwordspacing

\bibitem{sketchlearn}
\BIBentryALTinterwordspacing
Q.~Huang, P.~P.~C. Lee, and Y.~Bao, ``Sketchlearn: Relieving user burdens in
  approximate measurement with automated statistical inference,'' in
  \emph{Proceedings of the 2018 Conference of the ACM Special Interest Group on
  Data Communication}, ser. SIGCOMM '18.\hskip 1em plus 0.5em minus 0.4em\relax
  New York, NY, USA: ACM, 2018, pp. 576--590. [Online]. Available:
  \url{http://doi.acm.org/10.1145/3230543.3230559}
\BIBentrySTDinterwordspacing

\bibitem{nitro}
\BIBentryALTinterwordspacing
Z.~Liu, R.~Ben-Basat, G.~Einziger, Y.~Kassner, V.~Braverman, R.~Friedman, and
  V.~Sekar, ``Nitrosketch: Robust and general sketch-based monitoring in
  software switches,'' in \emph{Proceedings of the ACM Special Interest Group
  on Data Communication}, ser. SIGCOMM '19.\hskip 1em plus 0.5em minus
  0.4em\relax New York, NY, USA: Association for Computing Machinery, 2019, pp.
  334--350. [Online]. Available: \url{https://doi.org/10.1145/3341302.3342076}
\BIBentrySTDinterwordspacing

\bibitem{coco}
\BIBentryALTinterwordspacing
Y.~Zhang, Z.~Liu, R.~Wang, T.~Yang, J.~Li, R.~Miao, P.~Liu, R.~Zhang, and
  J.~Jiang, ``Cocosketch: High-performance sketch-based measurement over
  arbitrary partial key query,'' in \emph{Proceedings of the 2021 ACM SIGCOMM
  2021 Conference}, ser. SIGCOMM '21.\hskip 1em plus 0.5em minus 0.4em\relax
  New York, NY, USA: Association for Computing Machinery, 2021, pp. 207--222.
  [Online]. Available: \url{https://doi.org/10.1145/3452296.3472892}
\BIBentrySTDinterwordspacing

\bibitem{hashpipe}
\BIBentryALTinterwordspacing
V.~Sivaraman, S.~Narayana, O.~Rottenstreich, S.~Muthukrishnan, and J.~Rexford,
  ``Heavy-hitter detection entirely in the data plane,'' in \emph{Proceedings
  of the Symposium on SDN Research}, ser. SOSR '17.\hskip 1em plus 0.5em minus
  0.4em\relax New York, NY, USA: ACM, 2017, pp. 164--176. [Online]. Available:
  \url{http://doi.acm.org/10.1145/3050220.3063772}
\BIBentrySTDinterwordspacing

\bibitem{diana-sosr}
\BIBentryALTinterwordspacing
D.~A. Popescu, G.~Antichi, and A.~W. Moore, ``Enabling fast hierarchical heavy
  hitter detection using programmable data planes,'' in \emph{Proceedings of
  the Symposium on SDN Research}, ser. SOSR '17.\hskip 1em plus 0.5em minus
  0.4em\relax New York, NY, USA: ACM, 2017, pp. 191--192. [Online]. Available:
  \url{http://doi.acm.org/10.1145/3050220.3060606}
\BIBentrySTDinterwordspacing

\bibitem{diana-sosr2}
\BIBentryALTinterwordspacing
J.~Ku\v{c}era, D.~A. Popescu, H.~Wang, A.~Moore, J.~Ko\v{r}enek, and
  G.~Antichi, ``Enabling event-triggered data plane monitoring,'' in
  \emph{Proceedings of the Symposium on SDN Research}, ser. SOSR '20.\hskip 1em
  plus 0.5em minus 0.4em\relax New York, NY, USA: Association for Computing
  Machinery, 2020, pp. 14--26. [Online]. Available:
  \url{https://doi.org/10.1145/3373360.3380830}
\BIBentrySTDinterwordspacing

\bibitem{flowradar}
\BIBentryALTinterwordspacing
Y.~Li, R.~Miao, C.~Kim, and M.~Yu, ``Flowradar: A better netflow for data
  centers,'' in \emph{Proceedings of the 13th Usenix Conference on Networked
  Systems Design and Implementation}, ser. NSDI'16.\hskip 1em plus 0.5em minus
  0.4em\relax Berkeley, CA, USA: USENIX Association, 2016, pp. 311--324.
  [Online]. Available: \url{http://dl.acm.org/citation.cfm?id=2930611.2930632}
\BIBentrySTDinterwordspacing

\bibitem{marple}
\BIBentryALTinterwordspacing
S.~Narayana, A.~Sivaraman, V.~Nathan, P.~Goyal, V.~Arun, M.~Alizadeh,
  V.~Jeyakumar, and C.~Kim, ``Language-directed hardware design for network
  performance monitoring,'' in \emph{Proceedings of the Conference of the ACM
  Special Interest Group on Data Communication}, ser. SIGCOMM '17.\hskip 1em
  plus 0.5em minus 0.4em\relax New York, NY, USA: ACM, 2017, pp. 85--98.
  [Online]. Available: \url{http://doi.acm.org/10.1145/3098822.3098829}
\BIBentrySTDinterwordspacing

\bibitem{lossradar}
\BIBentryALTinterwordspacing
Y.~Li, R.~Miao, C.~Kim, and M.~Yu, ``Lossradar: Fast detection of lost packets
  in data center networks,'' in \emph{Proceedings of the 12th International on
  Conference on Emerging Networking EXperiments and Technologies}, ser. CoNEXT
  '16.\hskip 1em plus 0.5em minus 0.4em\relax New York, NY, USA: ACM, 2016, pp.
  481--495. [Online]. Available:
  \url{http://doi.acm.org/10.1145/2999572.2999609}
\BIBentrySTDinterwordspacing

\bibitem{int}
M.~Hira and L.~Wobker, ``{Improving Network Monitoring and Management with
  Programmable Data Planes},''
  \url{http://p4.org/p4/inband-network-telemetry/}, 2016, [Online; accessed
  December 2018].

\bibitem{pint}
\BIBentryALTinterwordspacing
R.~Ben~Basat, S.~Ramanathan, Y.~Li, G.~Antichi, M.~Yu, and M.~Mitzenmacher,
  ``Pint: Probabilistic in-band network telemetry,'' in \emph{Proceedings of
  the Annual Conference of the ACM Special Interest Group on Data Communication
  on the Applications, Technologies, Architectures, and Protocols for Computer
  Communication}, ser. SIGCOMM '20.\hskip 1em plus 0.5em minus 0.4em\relax New
  York, NY, USA: Association for Computing Machinery, 2020, pp. 662--680.
  [Online]. Available: \url{https://doi.org/10.1145/3387514.3405894}
\BIBentrySTDinterwordspacing

\bibitem{arzani2016}
\BIBentryALTinterwordspacing
B.~Arzani, S.~Ciraci, B.~T. Loo, A.~Schuster, and G.~Outhred, ``{Taking the
  Blame Game out of Data Centers Operations with NetPoirot},'' in
  \emph{Proceedings of the 2016 ACM SIGCOMM Conference}, ser. SIGCOMM
  '16.\hskip 1em plus 0.5em minus 0.4em\relax New York, NY, USA: ACM, 2016, pp.
  440--453. [Online]. Available:
  \url{http://doi.acm.org/10.1145/2934872.2934884}
\BIBentrySTDinterwordspacing

\bibitem{arjun2017}
\BIBentryALTinterwordspacing
A.~Roy, H.~Zeng, J.~Bagga, and A.~C. Snoeren, ``{Passive Realtime Datacenter
  Fault Detection and Localization},'' in \emph{14th {USENIX} Symposium on
  Networked Systems Design and Implementation ({NSDI} 17)}.\hskip 1em plus
  0.5em minus 0.4em\relax Boston, MA: {USENIX} Association, 2017, pp. 595--612.
  [Online]. Available:
  \url{https://www.usenix.org/conference/nsdi17/technical-sessions/presentation/roy}
\BIBentrySTDinterwordspacing

\bibitem{007}
\BIBentryALTinterwordspacing
B.~Arzani, S.~Ciraci, L.~Chamon, Y.~Zhu, H.~H. Liu, J.~Padhye, B.~T. Loo, and
  G.~Outhred, ``007: Democratically finding the cause of packet drops,'' in
  \emph{15th {USENIX} Symposium on Networked Systems Design and Implementation
  ({NSDI} 18)}.\hskip 1em plus 0.5em minus 0.4em\relax Renton, WA: {USENIX}
  Association, 2018, pp. 419--435. [Online]. Available:
  \url{https://www.usenix.org/conference/nsdi18/presentation/arzani}
\BIBentrySTDinterwordspacing

\bibitem{pathdump}
\BIBentryALTinterwordspacing
P.~Tammana, R.~Agarwal, and M.~Lee, ``Simplifying datacenter network debugging
  with pathdump,'' in \emph{12th {USENIX} Symposium on Operating Systems Design
  and Implementation ({OSDI} 16)}.\hskip 1em plus 0.5em minus 0.4em\relax
  Savannah, GA: {USENIX} Association, 2016, pp. 233--248. [Online]. Available:
  \url{https://www.usenix.org/conference/osdi16/technical-sessions/presentation/tammana}
\BIBentrySTDinterwordspacing

\bibitem{zhang2017}
\BIBentryALTinterwordspacing
Q.~Zhang, V.~Liu, H.~Zeng, and A.~Krishnamurthy, ``High-resolution measurement
  of data center microbursts,'' in \emph{Proceedings of the 2017 Internet
  Measurement Conference}, ser. IMC '17.\hskip 1em plus 0.5em minus 0.4em\relax
  New York, NY, USA: ACM, 2017, pp. 78--85. [Online]. Available:
  \url{http://doi.acm.org/10.1145/3131365.3131375}
\BIBentrySTDinterwordspacing

\bibitem{bosshart}
\BIBentryALTinterwordspacing
P.~Bosshart, G.~Gibb, H.-S. Kim, G.~Varghese, N.~McKeown, M.~Izzard, F.~Mujica,
  and M.~Horowitz, ``Forwarding metamorphosis: Fast programmable match-action
  processing in hardware for sdn,'' in \emph{Proceedings of the ACM SIGCOMM
  2013 Conference on SIGCOMM}, ser. SIGCOMM '13.\hskip 1em plus 0.5em minus
  0.4em\relax New York, NY, USA: ACM, 2013, pp. 99--110. [Online]. Available:
  \url{http://doi.acm.org/10.1145/2486001.2486011}
\BIBentrySTDinterwordspacing

\bibitem{sonata}
\BIBentryALTinterwordspacing
A.~Gupta, R.~Harrison, M.~Canini, N.~Feamster, J.~Rexford, and W.~Willinger,
  ``Sonata: Query-driven streaming network telemetry,'' in \emph{Proceedings of
  the 2018 Conference of the ACM Special Interest Group on Data Communication},
  ser. SIGCOMM '18.\hskip 1em plus 0.5em minus 0.4em\relax New York, NY, USA:
  ACM, 2018, pp. 357--371. [Online]. Available:
  \url{http://doi.acm.org/10.1145/3230543.3230555}
\BIBentrySTDinterwordspacing

\bibitem{gog}
I.~C. Gog, ``Flexible and efficient computation in large data centres,'' Ph.D.
  dissertation, University of Cambridge, 2017.

\bibitem{malte}
M.~Schwarzkopf, ``Operating system support for warehouse-scale computing,''
  Ph.D. dissertation, University of Cambridge, 2016.

\bibitem{bistro}
\BIBentryALTinterwordspacing
A.~Goder, A.~Spiridonov, and Y.~Wang, ``Bistro: Scheduling data-parallel jobs
  against live production systems,'' in \emph{2015 {USENIX} Annual Technical
  Conference ({USENIX} {ATC} 15)}.\hskip 1em plus 0.5em minus 0.4em\relax Santa
  Clara, CA: {USENIX} Association, 2015, pp. 459--471. [Online]. Available:
  \url{https://www.usenix.org/conference/atc15/technical-session/presentation/goder}
\BIBentrySTDinterwordspacing

\bibitem{quincy}
\BIBentryALTinterwordspacing
M.~Isard, V.~Prabhakaran, J.~Currey, U.~Wieder, K.~Talwar, and A.~Goldberg,
  ``Quincy: Fair scheduling for distributed computing clusters,'' in
  \emph{Proceedings of the ACM SIGOPS 22Nd Symposium on Operating Systems
  Principles}, ser. SOSP '09.\hskip 1em plus 0.5em minus 0.4em\relax New York,
  NY, USA: ACM, 2009, pp. 261--276. [Online]. Available:
  \url{http://doi.acm.org/10.1145/1629575.1629601}
\BIBentrySTDinterwordspacing

\bibitem{firmament}
\BIBentryALTinterwordspacing
I.~Gog, M.~Schwarzkopf, A.~Gleave, R.~N.~M. Watson, and S.~Hand, ``Firmament:
  Fast, centralized cluster scheduling at scale,'' in \emph{Proceedings of the
  12th USENIX Conference on Operating Systems Design and Implementation}, ser.
  OSDI'16.\hskip 1em plus 0.5em minus 0.4em\relax Berkeley, CA, USA: USENIX
  Association, 2016, pp. 99--115. [Online]. Available:
  \url{http://dl.acm.org/citation.cfm?id=3026877.3026886}
\BIBentrySTDinterwordspacing

\bibitem{tetrisched}
\BIBentryALTinterwordspacing
A.~Tumanov, T.~Zhu, J.~W. Park, M.~A. Kozuch, M.~Harchol-Balter, and G.~R.
  Ganger, ``Tetrisched: Global rescheduling with adaptive plan-ahead in dynamic
  heterogeneous clusters,'' in \emph{Proceedings of the Eleventh European
  Conference on Computer Systems}, ser. EuroSys '16.\hskip 1em plus 0.5em minus
  0.4em\relax New York, NY, USA: ACM, 2016, pp. 35:1--35:16. [Online].
  Available: \url{http://doi.acm.org/10.1145/2901318.2901355}
\BIBentrySTDinterwordspacing

\bibitem{apollo}
\BIBentryALTinterwordspacing
E.~Boutin, J.~Ekanayake, W.~Lin, B.~Shi, J.~Zhou, Z.~Qian, M.~Wu, and L.~Zhou,
  ``Apollo: Scalable and coordinated scheduling for cloud-scale computing,'' in
  \emph{11th {USENIX} Symposium on Operating Systems Design and Implementation
  ({OSDI} 14)}.\hskip 1em plus 0.5em minus 0.4em\relax Broomfield, CO: {USENIX}
  Association, 2014, pp. 285--300. [Online]. Available:
  \url{https://www.usenix.org/conference/osdi14/technical-sessions/presentation/boutin}
\BIBentrySTDinterwordspacing

\bibitem{sparrow}
\BIBentryALTinterwordspacing
K.~Ousterhout, P.~Wendell, M.~Zaharia, and I.~Stoica, ``Sparrow: Distributed,
  low latency scheduling,'' in \emph{Proceedings of the Twenty-Fourth ACM
  Symposium on Operating Systems Principles}, ser. SOSP '13.\hskip 1em plus
  0.5em minus 0.4em\relax New York, NY, USA: ACM, 2013, pp. 69--84. [Online].
  Available: \url{http://doi.acm.org/10.1145/2517349.2522716}
\BIBentrySTDinterwordspacing

\bibitem{hawk}
\BIBentryALTinterwordspacing
P.~Delgado, F.~Dinu, A.-M. Kermarrec, and W.~Zwaenepoel, ``Hawk: Hybrid
  datacenter scheduling,'' in \emph{Proceedings of the 2015 USENIX Conference
  on Usenix Annual Technical Conference}, ser. USENIX ATC '15.\hskip 1em plus
  0.5em minus 0.4em\relax Berkeley, CA, USA: USENIX Association, 2015, pp.
  499--510. [Online]. Available:
  \url{http://dl.acm.org/citation.cfm?id=2813767.2813804}
\BIBentrySTDinterwordspacing

\bibitem{eagle}
\BIBentryALTinterwordspacing
P.~Delgado, D.~Didona, F.~Dinu, and W.~Zwaenepoel, ``Job-aware scheduling in
  eagle: Divide and stick to your probes,'' in \emph{Proceedings of the Seventh
  ACM Symposium on Cloud Computing}, ser. SoCC '16.\hskip 1em plus 0.5em minus
  0.4em\relax New York, NY, USA: ACM, 2016, pp. 497--509. [Online]. Available:
  \url{http://doi.acm.org/10.1145/2987550.2987563}
\BIBentrySTDinterwordspacing

\bibitem{mercury}
\BIBentryALTinterwordspacing
K.~Karanasos, S.~Rao, C.~Curino, C.~Douglas, K.~Chaliparambil, G.~M. Fumarola,
  S.~Heddaya, R.~Ramakrishnan, and S.~Sakalanaga, ``Mercury: Hybrid centralized
  and distributed scheduling in large shared clusters,'' in \emph{2015 {USENIX}
  Annual Technical Conference ({USENIX} {ATC} 15)}.\hskip 1em plus 0.5em minus
  0.4em\relax Santa Clara, CA: {USENIX} Association, 2015, pp. 485--497.
  [Online]. Available:
  \url{https://www.usenix.org/conference/atc15/technical-session/presentation/karanasos}
\BIBentrySTDinterwordspacing

\bibitem{mesos}
\BIBentryALTinterwordspacing
B.~Hindman, A.~Konwinski, M.~Zaharia, A.~Ghodsi, A.~D. Joseph, R.~Katz,
  S.~Shenker, and I.~Stoica, ``Mesos: A platform for fine-grained resource
  sharing in the data center,'' in \emph{Proceedings of the 8th USENIX
  Conference on Networked Systems Design and Implementation}, ser.
  NSDI'11.\hskip 1em plus 0.5em minus 0.4em\relax Berkeley, CA, USA: USENIX
  Association, 2011, pp. 295--308. [Online]. Available:
  \url{http://dl.acm.org/citation.cfm?id=1972457.1972488}
\BIBentrySTDinterwordspacing

\bibitem{yarn}
\BIBentryALTinterwordspacing
V.~K. Vavilapalli, A.~C. Murthy, C.~Douglas, S.~Agarwal, M.~Konar, R.~Evans,
  T.~Graves, J.~Lowe, H.~Shah, S.~Seth, B.~Saha, C.~Curino, O.~O'Malley,
  S.~Radia, B.~Reed, and E.~Baldeschwieler, ``Apache hadoop yarn: Yet another
  resource negotiator,'' in \emph{Proceedings of the 4th Annual Symposium on
  Cloud Computing}, ser. SOCC '13.\hskip 1em plus 0.5em minus 0.4em\relax New
  York, NY, USA: ACM, 2013, pp. 5:1--5:16. [Online]. Available:
  \url{http://doi.acm.org/10.1145/2523616.2523633}
\BIBentrySTDinterwordspacing

\bibitem{jockey}
\BIBentryALTinterwordspacing
A.~D. Ferguson, P.~Bodik, S.~Kandula, E.~Boutin, and R.~Fonseca, ``Jockey:
  Guaranteed job latency in data parallel clusters,'' in \emph{Proceedings of
  the 7th ACM European Conference on Computer Systems}, ser. EuroSys '12.\hskip
  1em plus 0.5em minus 0.4em\relax New York, NY, USA: ACM, 2012, pp. 99--112.
  [Online]. Available: \url{http://doi.acm.org/10.1145/2168836.2168847}
\BIBentrySTDinterwordspacing

\bibitem{omega}
\BIBentryALTinterwordspacing
M.~Schwarzkopf, A.~Konwinski, M.~Abd-El-Malek, and J.~Wilkes, ``Omega:
  Flexible, scalable schedulers for large compute clusters,'' in
  \emph{Proceedings of the 8th ACM European Conference on Computer Systems},
  ser. EuroSys '13.\hskip 1em plus 0.5em minus 0.4em\relax New York, NY, USA:
  ACM, 2013, pp. 351--364. [Online]. Available:
  \url{http://doi.acm.org/10.1145/2465351.2465386}
\BIBentrySTDinterwordspacing

\bibitem{quasar}
\BIBentryALTinterwordspacing
C.~Delimitrou and C.~Kozyrakis, ``Quasar: Resource-efficient and qos-aware
  cluster management,'' in \emph{Proceedings of the 19th International
  Conference on Architectural Support for Programming Languages and Operating
  Systems}, ser. ASPLOS '14.\hskip 1em plus 0.5em minus 0.4em\relax New York,
  NY, USA: ACM, 2014, pp. 127--144. [Online]. Available:
  \url{http://doi.acm.org/10.1145/2541940.2541941}
\BIBentrySTDinterwordspacing

\bibitem{tetris}
\BIBentryALTinterwordspacing
R.~Grandl, G.~Ananthanarayanan, S.~Kandula, S.~Rao, and A.~Akella,
  ``Multi-resource packing for cluster schedulers,'' in \emph{Proceedings of
  the 2014 ACM Conference on SIGCOMM}, ser. SIGCOMM '14.\hskip 1em plus 0.5em
  minus 0.4em\relax New York, NY, USA: ACM, 2014, pp. 455--466. [Online].
  Available: \url{http://doi.acm.org/10.1145/2619239.2626334}
\BIBentrySTDinterwordspacing

\bibitem{altruistic}
\BIBentryALTinterwordspacing
R.~Grandl, M.~Chowdhury, A.~Akella, and G.~Ananthanarayanan, ``Altruistic
  scheduling in multi-resource clusters,'' in \emph{Proceedings of the 12th
  USENIX Conference on Operating Systems Design and Implementation}, ser.
  OSDI'16.\hskip 1em plus 0.5em minus 0.4em\relax Berkeley, CA, USA: USENIX
  Association, 2016, pp. 65--80. [Online]. Available:
  \url{http://dl.acm.org/citation.cfm?id=3026877.3026884}
\BIBentrySTDinterwordspacing

\bibitem{morpheus}
\BIBentryALTinterwordspacing
S.~A. Jyothi, C.~Curino, I.~Menache, S.~M. Narayanamurthy, A.~Tumanov,
  J.~Yaniv, R.~Mavlyutov, I.~n. Goiri, S.~Krishnan, J.~Kulkarni, and S.~Rao,
  ``Morpheus: Towards automated slos for enterprise clusters,'' in
  \emph{Proceedings of the 12th USENIX Conference on Operating Systems Design
  and Implementation}, ser. OSDI'16.\hskip 1em plus 0.5em minus 0.4em\relax
  Berkeley, CA, USA: USENIX Association, 2016, pp. 117--134. [Online].
  Available: \url{http://dl.acm.org/citation.cfm?id=3026877.3026887}
\BIBentrySTDinterwordspacing

\bibitem{ernest}
\BIBentryALTinterwordspacing
S.~Venkataraman, Z.~Yang, M.~Franklin, B.~Recht, and I.~Stoica, ``Ernest:
  Efficient performance prediction for large-scale advanced analytics,'' in
  \emph{Proceedings of the 13th Usenix Conference on Networked Systems Design
  and Implementation}, ser. NSDI'16.\hskip 1em plus 0.5em minus 0.4em\relax
  Berkeley, CA, USA: USENIX Association, 2016, pp. 363--378. [Online].
  Available: \url{http://dl.acm.org/citation.cfm?id=2930611.2930635}
\BIBentrySTDinterwordspacing

\bibitem{cherrypick}
\BIBentryALTinterwordspacing
O.~Alipourfard, H.~H. Liu, J.~Chen, S.~Venkataraman, M.~Yu, and M.~Zhang,
  ``Cherrypick: Adaptively unearthing the best cloud configurations for big
  data analytics,'' in \emph{Proceedings of the 14th USENIX Conference on
  Networked Systems Design and Implementation}, ser. NSDI'17.\hskip 1em plus
  0.5em minus 0.4em\relax Berkeley, CA, USA: USENIX Association, 2017, pp.
  469--482. [Online]. Available:
  \url{http://dl.acm.org/citation.cfm?id=3154630.3154669}
\BIBentrySTDinterwordspacing

\bibitem{optimus}
\BIBentryALTinterwordspacing
Y.~Peng, Y.~Bao, Y.~Chen, C.~Wu, and C.~Guo, ``Optimus: An efficient dynamic
  resource scheduler for deep learning clusters,'' in \emph{Proceedings of the
  Thirteenth EuroSys Conference}, ser. EuroSys '18.\hskip 1em plus 0.5em minus
  0.4em\relax New York, NY, USA: ACM, 2018, pp. 3:1--3:14. [Online]. Available:
  \url{http://doi.acm.org/10.1145/3190508.3190517}
\BIBentrySTDinterwordspacing

\bibitem{paragon}
\BIBentryALTinterwordspacing
C.~Delimitrou and C.~Kozyrakis, ``Paragon: Qos-aware scheduling for
  heterogeneous datacenters,'' in \emph{Proceedings of the Eighteenth
  International Conference on Architectural Support for Programming Languages
  and Operating Systems}, ser. ASPLOS '13.\hskip 1em plus 0.5em minus
  0.4em\relax New York, NY, USA: ACM, 2013, pp. 77--88. [Online]. Available:
  \url{http://doi.acm.org/10.1145/2451116.2451125}
\BIBentrySTDinterwordspacing

\bibitem{tpu}
\BIBentryALTinterwordspacing
N.~P. Jouppi, C.~Young, N.~Patil, D.~Patterson, G.~Agrawal, R.~Bajwa, S.~Bates,
  S.~Bhatia, N.~Boden, A.~Borchers, R.~Boyle, P.-l. Cantin, C.~Chao, C.~Clark,
  J.~Coriell, M.~Daley, M.~Dau, J.~Dean, B.~Gelb, T.~V. Ghaemmaghami,
  R.~Gottipati, W.~Gulland, R.~Hagmann, C.~R. Ho, D.~Hogberg, J.~Hu, R.~Hundt,
  D.~Hurt, J.~Ibarz, A.~Jaffey, A.~Jaworski, A.~Kaplan, H.~Khaitan,
  D.~Killebrew, A.~Koch, N.~Kumar, S.~Lacy, J.~Laudon, J.~Law, D.~Le, C.~Leary,
  Z.~Liu, K.~Lucke, A.~Lundin, G.~MacKean, A.~Maggiore, M.~Mahony, K.~Miller,
  R.~Nagarajan, R.~Narayanaswami, R.~Ni, K.~Nix, T.~Norrie, M.~Omernick,
  N.~Penukonda, A.~Phelps, J.~Ross, M.~Ross, A.~Salek, E.~Samadiani, C.~Severn,
  G.~Sizikov, M.~Snelham, J.~Souter, D.~Steinberg, A.~Swing, M.~Tan,
  G.~Thorson, B.~Tian, H.~Toma, E.~Tuttle, V.~Vasudevan, R.~Walter, W.~Wang,
  E.~Wilcox, and D.~H. Yoon, ``In-datacenter performance analysis of a tensor
  processing unit,'' in \emph{Proceedings of the 44th Annual International
  Symposium on Computer Architecture}, ser. ISCA '17.\hskip 1em plus 0.5em
  minus 0.4em\relax New York, NY, USA: ACM, 2017, pp. 1--12. [Online].
  Available: \url{http://doi.acm.org/10.1145/3079856.3080246}
\BIBentrySTDinterwordspacing

\bibitem{sharma:2011}
\BIBentryALTinterwordspacing
B.~Sharma, V.~Chudnovsky, J.~L. Hellerstein, R.~Rifaat, and C.~R. Das,
  ``Modeling and synthesizing task placement constraints in google compute
  clusters,'' in \emph{Proceedings of the 2Nd ACM Symposium on Cloud
  Computing}, ser. SOCC '11.\hskip 1em plus 0.5em minus 0.4em\relax New York,
  NY, USA: ACM, 2011, pp. 3:1--3:14. [Online]. Available:
  \url{http://doi.acm.org/10.1145/2038916.2038919}
\BIBentrySTDinterwordspacing

\bibitem{medea}
\BIBentryALTinterwordspacing
P.~Garefalakis, K.~Karanasos, P.~Pietzuch, A.~Suresh, and S.~Rao, ``Medea:
  Scheduling of long running applications in shared production clusters,'' in
  \emph{Proceedings of the Thirteenth EuroSys Conference}, ser. EuroSys
  '18.\hskip 1em plus 0.5em minus 0.4em\relax New York, NY, USA: ACM, 2018, pp.
  4:1--4:13. [Online]. Available:
  \url{http://doi.acm.org/10.1145/3190508.3190549}
\BIBentrySTDinterwordspacing

\bibitem{kubernetes}
K.~Hightower, B.~Burns, and J.~Beda, \emph{Kubernetes: Up and Running Dive into
  the Future of Infrastructure}, 1st~ed.\hskip 1em plus 0.5em minus 0.4em\relax
  O'Reilly Media, Inc., 2017.

\bibitem{corral}
\BIBentryALTinterwordspacing
V.~Jalaparti, P.~Bodik, I.~Menache, S.~Rao, K.~Makarychev, and M.~Caesar,
  ``Network-aware scheduling for data-parallel jobs: Plan when you can,'' in
  \emph{Proceedings of the 2015 ACM Conference on Special Interest Group on
  Data Communication}, ser. SIGCOMM '15.\hskip 1em plus 0.5em minus 0.4em\relax
  New York, NY, USA: ACM, 2015, pp. 407--420. [Online]. Available:
  \url{http://doi.acm.org/10.1145/2785956.2787488}
\BIBentrySTDinterwordspacing

\bibitem{ballani2011}
\BIBentryALTinterwordspacing
H.~Ballani, P.~Costa, T.~Karagiannis, and A.~Rowstron, ``Towards predictable
  datacenter networks,'' in \emph{Proceedings of the ACM SIGCOMM 2011
  Conference}, ser. SIGCOMM '11.\hskip 1em plus 0.5em minus 0.4em\relax New
  York, NY, USA: ACM, 2011, pp. 242--253. [Online]. Available:
  \url{http://doi.acm.org/10.1145/2018436.2018465}
\BIBentrySTDinterwordspacing

\bibitem{persico-ec2}
\BIBentryALTinterwordspacing
V.~Persico, P.~Marchetta, A.~Botta, and A.~Pescap\`{e}, ``Measuring network
  throughput in the cloud,'' \emph{Comput. Netw.}, vol.~93, no.~P3, pp.
  408--422, Dec. 2015. [Online]. Available:
  \url{http://dx.doi.org/10.1016/j.comnet.2015.09.037}
\BIBentrySTDinterwordspacing

\bibitem{persico-azure}
V.~Persico, P.~Marchetta, A.~Botta, and A.~Pescape, ``On network throughput
  variability in microsoft azure cloud,'' in \emph{2015 IEEE Global
  Communications Conference (GLOBECOM)}, Dec 2015, pp. 1--6.

\bibitem{hose}
\BIBentryALTinterwordspacing
N.~G. Duffield, P.~Goyal, A.~Greenberg, P.~Mishra, K.~K. Ramakrishnan, and
  J.~E. van~der Merive, ``A flexible model for resource management in virtual
  private networks,'' in \emph{Proceedings of the Conference on Applications,
  Technologies, Architectures, and Protocols for Computer Communication}, ser.
  SIGCOMM '99.\hskip 1em plus 0.5em minus 0.4em\relax New York, NY, USA: ACM,
  1999, pp. 95--108. [Online]. Available:
  \url{http://doi.acm.org/10.1145/316188.316209}
\BIBentrySTDinterwordspacing

\bibitem{secondnet}
\BIBentryALTinterwordspacing
C.~Guo, G.~Lu, H.~J. Wang, S.~Yang, C.~Kong, P.~Sun, W.~Wu, and Y.~Zhang,
  ``Secondnet: A data center network virtualization architecture with bandwidth
  guarantees,'' in \emph{Proceedings of the 6th International COnference}, ser.
  Co-NEXT '10.\hskip 1em plus 0.5em minus 0.4em\relax New York, NY, USA: ACM,
  2010, pp. 15:1--15:12. [Online]. Available:
  \url{http://doi.acm.org/10.1145/1921168.1921188}
\BIBentrySTDinterwordspacing

\bibitem{gatekeeper}
\BIBentryALTinterwordspacing
H.~Rodrigues, J.~R. Santos, Y.~Turner, P.~Soares, and D.~Guedes, ``Gatekeeper:
  Supporting bandwidth guarantees for multi-tenant datacenter networks,'' in
  \emph{Proceedings of the 3rd Conference on I/O Virtualization}, ser.
  WIOV'11.\hskip 1em plus 0.5em minus 0.4em\relax Berkeley, CA, USA: USENIX
  Association, 2011, pp. 6--6. [Online]. Available:
  \url{http://dl.acm.org/citation.cfm?id=2001555.2001561}
\BIBentrySTDinterwordspacing

\bibitem{eyeq}
\BIBentryALTinterwordspacing
V.~Jeyakumar, M.~Alizadeh, D.~Mazi\`{e}res, B.~Prabhakar, C.~Kim, and
  A.~Greenberg, ``{EyeQ: Practical Network Performance Isolation at the
  Edge},'' in \emph{Proceedings of the 10th USENIX Conference on Networked
  Systems Design and Implementation}, ser. nsdi'13.\hskip 1em plus 0.5em minus
  0.4em\relax Berkeley, CA, USA: USENIX Association, 2013, pp. 297--312.
  [Online]. Available: \url{http://dl.acm.org/citation.cfm?id=2482626.2482656}
\BIBentrySTDinterwordspacing

\bibitem{elastic}
\BIBentryALTinterwordspacing
L.~Popa, P.~Yalagandula, S.~Banerjee, J.~C. Mogul, Y.~Turner, and J.~R. Santos,
  ``Elasticswitch: Practical work-conserving bandwidth guarantees for cloud
  computing,'' in \emph{Proceedings of the ACM SIGCOMM 2013 Conference on
  SIGCOMM}, ser. SIGCOMM '13.\hskip 1em plus 0.5em minus 0.4em\relax New York,
  NY, USA: ACM, 2013, pp. 351--362. [Online]. Available:
  \url{http://doi.acm.org/10.1145/2486001.2486027}
\BIBentrySTDinterwordspacing

\bibitem{proteus}
\BIBentryALTinterwordspacing
D.~Xie, N.~Ding, Y.~C. Hu, and R.~Kompella, ``The only constant is change:
  Incorporating time-varying network reservations in data centers,'' in
  \emph{Proceedings of the ACM SIGCOMM 2012 Conference on Applications,
  Technologies, Architectures, and Protocols for Computer Communication}, ser.
  SIGCOMM '12.\hskip 1em plus 0.5em minus 0.4em\relax New York, NY, USA: ACM,
  2012, pp. 199--210. [Online]. Available:
  \url{http://doi.acm.org/10.1145/2342356.2342397}
\BIBentrySTDinterwordspacing

\bibitem{cloudmirror}
\BIBentryALTinterwordspacing
J.~Lee, Y.~Turner, M.~Lee, L.~Popa, S.~Banerjee, J.-M. Kang, and P.~Sharma,
  ``Application-driven bandwidth guarantees in datacenters,'' in
  \emph{Proceedings of the 2014 ACM Conference on SIGCOMM}, ser. SIGCOMM
  '14.\hskip 1em plus 0.5em minus 0.4em\relax New York, NY, USA: ACM, 2014, pp.
  467--478. [Online]. Available:
  \url{http://doi.acm.org/10.1145/2619239.2626326}
\BIBentrySTDinterwordspacing

\bibitem{pulsar}
\BIBentryALTinterwordspacing
S.~Angel, H.~Ballani, T.~Karagiannis, G.~O'Shea, and E.~Thereska, ``End-to-end
  performance isolation through virtual datacenters,'' in \emph{Proceedings of
  the 11th USENIX Conference on Operating Systems Design and Implementation},
  ser. OSDI'14.\hskip 1em plus 0.5em minus 0.4em\relax Berkeley, CA, USA:
  USENIX Association, 2014, pp. 233--248. [Online]. Available:
  \url{http://dl.acm.org/citation.cfm?id=2685048.2685067}
\BIBentrySTDinterwordspacing

\bibitem{seawell}
\BIBentryALTinterwordspacing
A.~Shieh, S.~Kandula, A.~Greenberg, C.~Kim, and B.~Saha, ``Sharing the data
  center network,'' in \emph{Proceedings of the 8th USENIX Conference on
  Networked Systems Design and Implementation}, ser. NSDI'11.\hskip 1em plus
  0.5em minus 0.4em\relax Berkeley, CA, USA: USENIX Association, 2011, pp.
  309--322. [Online]. Available:
  \url{http://dl.acm.org/citation.cfm?id=1972457.1972489}
\BIBentrySTDinterwordspacing

\bibitem{netshare}
\BIBentryALTinterwordspacing
V.~T. Lam, S.~Radhakrishnan, R.~Pan, A.~Vahdat, and G.~Varghese, ``Netshare and
  stochastic netshare: Predictable bandwidth allocation for data centers,''
  \emph{SIGCOMM Comput. Commun. Rev.}, vol.~42, no.~3, pp. 5--11, Jun. 2012.
  [Online]. Available: \url{http://doi.acm.org/10.1145/2317307.2317309}
\BIBentrySTDinterwordspacing

\bibitem{faircloud}
\BIBentryALTinterwordspacing
L.~Popa, G.~Kumar, M.~Chowdhury, A.~Krishnamurthy, S.~Ratnasamy, and I.~Stoica,
  ``Faircloud: Sharing the network in cloud computing,'' in \emph{Proceedings
  of the ACM SIGCOMM 2012 Conference on Applications, Technologies,
  Architectures, and Protocols for Computer Communication}, ser. SIGCOMM
  '12.\hskip 1em plus 0.5em minus 0.4em\relax New York, NY, USA: ACM, 2012, pp.
  187--198. [Online]. Available:
  \url{http://doi.acm.org/10.1145/2342356.2342396}
\BIBentrySTDinterwordspacing

\bibitem{ballani2013}
\BIBentryALTinterwordspacing
H.~Ballani, K.~Jang, T.~Karagiannis, C.~Kim, D.~Gunawardena, and G.~O'Shea,
  ``Chatty tenants and the cloud network sharing problem,'' in
  \emph{Proceedings of the 10th USENIX Conference on Networked Systems Design
  and Implementation}, ser. nsdi'13.\hskip 1em plus 0.5em minus 0.4em\relax
  Berkeley, CA, USA: USENIX Association, 2013, pp. 171--184. [Online].
  Available: \url{http://dl.acm.org/citation.cfm?id=2482626.2482644}
\BIBentrySTDinterwordspacing

\bibitem{choreo}
\BIBentryALTinterwordspacing
K.~LaCurts, S.~Deng, A.~Goyal, and H.~Balakrishnan, ``Choreo: Network-aware
  task placement for cloud applications,'' in \emph{Proceedings of the 2013
  Conference on Internet Measurement Conference}, ser. IMC '13.\hskip 1em plus
  0.5em minus 0.4em\relax New York, NY, USA: ACM, 2013, pp. 191--204. [Online].
  Available: \url{http://doi.acm.org/10.1145/2504730.2504744}
\BIBentrySTDinterwordspacing

\bibitem{cicada}
\BIBentryALTinterwordspacing
K.~LaCurts, J.~C. Mogul, H.~Balakrishnan, and Y.~Turner, ``Cicada: Introducing
  predictive guarantees for cloud networks,'' in \emph{Proceedings of the 6th
  USENIX Conference on Hot Topics in Cloud Computing}, ser. HotCloud'14.\hskip
  1em plus 0.5em minus 0.4em\relax Berkeley, CA, USA: USENIX Association, 2014,
  pp. 14--14. [Online]. Available:
  \url{http://dl.acm.org/citation.cfm?id=2696535.2696549}
\BIBentrySTDinterwordspacing

\bibitem{silo}
\BIBentryALTinterwordspacing
K.~Jang, J.~Sherry, H.~Ballani, and T.~Moncaster, ``Silo: Predictable message
  latency in the cloud,'' in \emph{Proceedings of the 2015 ACM Conference on
  Special Interest Group on Data Communication}, ser. SIGCOMM '15.\hskip 1em
  plus 0.5em minus 0.4em\relax New York, NY, USA: ACM, 2015, pp. 435--448.
  [Online]. Available: \url{http://doi.acm.org/10.1145/2785956.2787479}
\BIBentrySTDinterwordspacing

\bibitem{sncmeister}
\BIBentryALTinterwordspacing
T.~Zhu, D.~S. Berger, and M.~Harchol-Balter, ``Snc-meister: Admitting more
  tenants with tail latency slos,'' in \emph{Proceedings of the Seventh ACM
  Symposium on Cloud Computing}, ser. SoCC '16.\hskip 1em plus 0.5em minus
  0.4em\relax New York, NY, USA: ACM, 2016, pp. 374--387. [Online]. Available:
  \url{http://doi.acm.org/10.1145/2987550.2987585}
\BIBentrySTDinterwordspacing

\bibitem{diana-nomora}
D.~A. Popescu and A.~W. Moore, ``{Network Latency and Application Performance
  Aware Cluster Scheduling in Data Centers},'' \emph{IEEE Network}, vol.~36,
  no.~2, pp. 58--65, 2022.

\bibitem{Dean:2013}
\BIBentryALTinterwordspacing
J.~Dean and L.~A. Barroso, ``The tail at scale,'' \emph{Commun. ACM}, vol.~56,
  no.~2, pp. 74--80, Feb. 2013. [Online]. Available:
  \url{http://doi.acm.org/10.1145/2408776.2408794}
\BIBentrySTDinterwordspacing

\bibitem{detail}
\BIBentryALTinterwordspacing
D.~Zats, T.~Das, P.~Mohan, D.~Borthakur, and R.~Katz, ``Detail: Reducing the
  flow completion time tail in datacenter networks,'' \emph{SIGCOMM Comput.
  Commun. Rev.}, vol.~42, no.~4, pp. 139--150, Aug. 2012. [Online]. Available:
  \url{http://doi.acm.org/10.1145/2377677.2377711}
\BIBentrySTDinterwordspacing

\bibitem{bobtail}
\BIBentryALTinterwordspacing
Y.~Xu, Z.~Musgrave, B.~Noble, and M.~Bailey, ``Bobtail: Avoiding long tails in
  the cloud,'' in \emph{Proceedings of the 10th USENIX Conference on Networked
  Systems Design and Implementation}, ser. nsdi'13.\hskip 1em plus 0.5em minus
  0.4em\relax Berkeley, CA, USA: USENIX Association, 2013, pp. 329--342.
  [Online]. Available: \url{http://dl.acm.org/citation.cfm?id=2482626.2482658}
\BIBentrySTDinterwordspacing

\bibitem{sigmetrics-moore}
\BIBentryALTinterwordspacing
A.~W. Moore and D.~Zuev, ``Internet traffic classification using bayesian
  analysis techniques,'' in \emph{Proceedings of the 2005 ACM SIGMETRICS
  International Conference on Measurement and Modeling of Computer Systems},
  ser. SIGMETRICS '05.\hskip 1em plus 0.5em minus 0.4em\relax New York, NY,
  USA: ACM, 2005, pp. 50--60. [Online]. Available:
  \url{http://doi.acm.org/10.1145/1064212.1064220}
\BIBentrySTDinterwordspacing

\bibitem{decima}
\BIBentryALTinterwordspacing
H.~Mao, M.~Schwarzkopf, S.~B. Venkatakrishnan, Z.~Meng, and M.~Alizadeh,
  ``{Learning Scheduling Algorithms for Data Processing Clusters},'' in
  \emph{Proceedings of the ACM Special Interest Group on Data Communication},
  ser. SIGCOMM '19.\hskip 1em plus 0.5em minus 0.4em\relax New York, NY, USA:
  Association for Computing Machinery, 2019, pp. 270--288. [Online]. Available:
  \url{https://doi.org/10.1145/3341302.3342080}
\BIBentrySTDinterwordspacing

\bibitem{graphene}
\BIBentryALTinterwordspacing
R.~Grandl, S.~Kandula, S.~Rao, A.~Akella, and J.~Kulkarni, ``{GRAPHENE}:
  Packing and dependency-aware scheduling for data-parallel clusters,'' in
  \emph{12th {USENIX} Symposium on Operating Systems Design and Implementation
  ({OSDI} 16)}.\hskip 1em plus 0.5em minus 0.4em\relax Savannah, GA: {USENIX}
  Association, Nov. 2016, pp. 81--97. [Online]. Available:
  \url{https://www.usenix.org/conference/osdi16/technical-sessions/presentation/grandl_graphene}
\BIBentrySTDinterwordspacing

\bibitem{deeprm}
\BIBentryALTinterwordspacing
H.~Mao, M.~Alizadeh, I.~Menache, and S.~Kandula, ``Resource management with
  deep reinforcement learning,'' in \emph{Proceedings of the 15th ACM Workshop
  on Hot Topics in Networks}, ser. HotNets '16.\hskip 1em plus 0.5em minus
  0.4em\relax New York, NY, USA: Association for Computing Machinery, 2016, pp.
  50--56. [Online]. Available: \url{https://doi.org/10.1145/3005745.3005750}
\BIBentrySTDinterwordspacing

\bibitem{rlscheduler}
D.~Zhang, D.~Dai, Y.~He, F.~S. Bao, and B.~Xie, \emph{RLScheduler: An Automated
  HPC Batch Job Scheduler Using Reinforcement Learning}.\hskip 1em plus 0.5em
  minus 0.4em\relax IEEE Press, 2020.

\bibitem{autopilot}
\BIBentryALTinterwordspacing
K.~Rzadca, P.~Findeisen, J.~Swiderski, P.~Zych, P.~Broniek, J.~Kusmierek,
  P.~Nowak, B.~Strack, P.~Witusowski, S.~Hand, and J.~Wilkes, ``Autopilot:
  Workload autoscaling at google,'' in \emph{Proceedings of the Fifteenth
  European Conference on Computer Systems}, ser. EuroSys '20.\hskip 1em plus
  0.5em minus 0.4em\relax New York, NY, USA: Association for Computing
  Machinery, 2020. [Online]. Available:
  \url{https://doi.org/10.1145/3342195.3387524}
\BIBentrySTDinterwordspacing

\bibitem{sosp-huawei}
\BIBentryALTinterwordspacing
S.~Bergsma, T.~Zeyl, A.~Senderovich, and J.~C. Beck, ``Generating complex,
  realistic cloud workloads using recurrent neural networks,'' in
  \emph{Proceedings of the ACM SIGOPS 28th Symposium on Operating Systems
  Principles}, ser. SOSP '21.\hskip 1em plus 0.5em minus 0.4em\relax New York,
  NY, USA: Association for Computing Machinery, 2021, pp. 376--391. [Online].
  Available: \url{https://doi.org/10.1145/3477132.3483590}
\BIBentrySTDinterwordspacing

\bibitem{sigmetrics-traces}
\BIBentryALTinterwordspacing
C.~Avin, M.~Ghobadi, C.~Griner, and S.~Schmid, ``On the complexity of traffic
  traces and implications,'' in \emph{Abstracts of the 2020
  SIGMETRICS/Performance Joint International Conference on Measurement and
  Modeling of Computer Systems}, ser. SIGMETRICS '20.\hskip 1em plus 0.5em
  minus 0.4em\relax New York, NY, USA: Association for Computing Machinery,
  2020, pp. 47--48. [Online]. Available:
  \url{https://doi.org/10.1145/3393691.3394205}
\BIBentrySTDinterwordspacing

\end{thebibliography}

%








\end{document}